\documentclass[manuscript]{aastex}
\usepackage{natbib}
\usepackage{enumerate}
\slugcomment{Not to appear in Nonlearned J., 45.}

\shorttitle{rotating outflow and the disk}
\shortauthors{Hara et al. }

\begin{document}


\title{The Rotating Outflow, Envelope and Disk \\in Class-0/I protostar [BHB2007] \#11 in the Pipe Nebula}

\author{C. Hara}
\affil{The University of Tokyo\altaffilmark{1} / National Astronomical Observatory of Japan\altaffilmark{2}}
\email{c.hara@nao.ac.jp}

\author{Y. Shimajiri}
\affil{Nobeyama Radio Observatory\altaffilmark{3} / National Astronomical Observatory of Japan\altaffilmark{2}}

\author{T. Tsukagoshi}
\affil{Ibaraki University\altaffilmark{4}}

\author{Y. Kurono, K. Saigo}
\affil{National Astronomical Observatory of Japan\altaffilmark{2}}

\author{F. Nakamura}
\affil{National Astronomical Observatory of Japan\altaffilmark{2}}

\author{M. Saito}
\affil{National Astronomical Observatory of Japan\altaffilmark{2} / Joint ALMA Observatory\altaffilmark{5}}

\author{David Wilner}
\affil{Harvard Smithsonian Center for Astrophysics\altaffilmark{6}}

\and

\author{R. Kawabe}
\affil{National Astronomical Observatory of Japan\altaffilmark{2} / Joint ALMA Observatory\altaffilmark{5}}

\altaffiltext{1}{7-3-1 Hongo Bunkyo, Tokyo 113-0033, Japan}
\altaffiltext{2}{2-21-1 Osawa Mitaka, Tokyo 181-0015, Japan}
\altaffiltext{3}{462-2 Nobeyama Minamimaki, Minamisaku District, Nagano Prefecture 384-1305, Japan}
\altaffiltext{4}{2-1-1 Bunkyo Mito, Ibaraki Prefecture 310-8512, Japan }
\altaffiltext{5}{Alonso de Cordova 3107 Vitacura, Satiago 763 0355, Chile}
\altaffiltext{6}{Center for Astrophysics, 60 Garden Street, Cambridge, MA 02138}


\begin{abstract}

We present the results of observations toward a low-mass Class-0/I protostar, [BHB2007]\#11 (afterwards B59\#11) at the nearby ($d$=130 pc) star forming region, Barnard 59 (B59) in the Pipe Nebula with {the} Atacama Submillimeter Telescope Experiment (ASTE) 10 m telescope ($\sim$22\arcsec\hspace{1ex}resolution) in CO(3--2), HCO$^+$, H$^{13}$CO$^+$(4--3), and 1.1 mm dust-continuum emissions. We also show Submillimeter Array (SMA) data in $^{12}$CO, $^{13}$CO, C$^{18}$O(2--1), and 1.3 mm dust-continuum emissions with $\sim$5\arcsec\hspace{1ex}resolution. From ASTE CO(3--2) observations, we found that B59\#11 is blowing a collimated outflow whose axis {lies} almost on the plane of the sky. The outflow {traces well} a cavity-like structure seen in the 1.1 mm dust-continuum emission. The results of SMA $^{13}$CO and C$^{18}$O(2--1) observations have revealed that a compact and elongated structure of dense gas is associated with B59\#11, which is oriented perpendicular to the outflow axis. There is a compact dust condensation with a size of 350$\times$180 AU seen in the SMA 1.3 mm continuum map, and the direction of its major axis is almost the same as that of the dense gas elongation. The distributions of $^{13}$CO and C$^{18}$O emission also show the velocity gradients along their major axes, which {are} considered to {arise} from the envelope/disk rotation. From the detailed analysis of the SMA data, we infer that B59\#11 is surrounded by a Keplerian disk with a size of {less than} 350 AU. In addition, the SMA CO(2--1) image shows a velocity gradient in the outflow along the same direction as that of the dense gas rotation. We suggest that this velocity gradient shows a rotation of the outflow.

\end{abstract}


\keywords{ stars: formation --- ISM: clouds --- ISM: radio continuum --- ISM: molecules --- ISM: individual ([BHB2007]\#11)}



\section{Introduction}
\label{sec:introduction}
Stars are formed through the gravitational collapse of a molecular cloud core, and during the collapse, the system undergoes an increasing density of 20 orders of magnitude. In the course of the gravitational collapse, a core is considered to spin up and eventually a Keplerian disk is formed around a protostar and grows in size during the main accretion phases. 

Previous interferometric observations in molecular lines have found evidences of Keplerian disks around protostars in main accretion phases (e.g., \citet{bri07,lom08,jor09}). 
These protostars, however, have somewhat high bolometric temperatures ($T_{\rm bol}$=238 K, 391 K, 351 K, and 310 K for L1489-IRS, Elias 29, IRS 63, and IRS 43, respectively) indicating that these are more evolved protostars than the Class-0 phase, and the initial conditions of disks have not been revealed yet. \citet{taka12} and \citet{lee10} also found Keplerian disks around Class-I protobinary systems in earlier evolutionary phases {(L1551NE; $T_{\rm bol}$=91K, HH111; $T_{\rm bol}$=78 K)}. There are, however, few samples of Keplerian disks observationally identified around protostars in the early phases.

The Barnard 59 (B59) is an irregularly shaped dark cloud sitting at the end of the Pipe Nebula. Here, we adopt a distance {to} B59 {of} 130$^{+13}_{-20}$ pc \citep{lom06}, which is most-commonly used in the previous studies of the Pipe Nebula. Although \citet{al07} proposed a distance of 145$\pm16$ pc, our adopted distance is consistent within uncertainties of their analyses. It should be noted that masses estimated in this paper have uncertainties of $\sim$20 \% {according to an} uncertainty of the distance of $\sim$10 \%. \citet{onishi99} carried out mapping observations toward the Pipe Nebula in CO(1--0) and C$^{18}$O(1--0) lines, and detected 14 C$^{18}$O dense cores. A CO(1--0) outflow {was} detected only at B59, suggesting that B59 is an only active star-forming region in the Pipe Nebula \citep{onishi99}. Spitzer observations have revealed 20 low-mass young stellar objects (YSOs), and {suggest} that the star formation efficiency {of the cluster} is $\sim$20\% \citep{bro07}. More detailed photometry at MIPS bands has revealed that there are 15 low-mass YSOs in the 0.3$\times$0.3 pc area \citep{for09}. The median stellar age of B59 has been estimated to be $2.6^{+4.1}_{-2.6}$ Myr\citep{cov10}. [BHB2007]\#11 (hereafter, B59\#11) is a deeply embedded low-mass protostar in the B59 region and classified as a Class-0/I object, which is in the transition phase from Class-0 to I with a bolometric temperature of 70 K \citep{bro07} and considered to be younger than the B59 median stellar age of $\sim$2.6 Myr. It is the strongest 70 \micron\hspace{1ex}emission source in the B59 region and has a bolometric luminosity of 2.2$\pm$0.3 $L_\sun$. \citet{bro07} detected extended IRAC 3.6 \micron \hspace{1ex}and 4.5 \micron \hspace{1ex}emissions on the northeast of B59\#11. These structures imply that a molecular outflow ejected from B59\#11 creates a cavity. \citet{riaz09} analyzed these extended nebulosities and have suggested that {the} inclination angle of the outflow ejected from B59\#11 is 53\arcdeg-59\arcdeg. \citet{dua12}, however, found that the outflow associated with B59\#11 is ejected almost in the plane of the sky from observations of molecular outflows in the CO(3--2) line. \citet{riaz09} also pointed out that B59\#11 is building up a weakly bounded binary system with 2M17112255-27243448 (hereafter, B59\#11SW; the apparent separation is $\sim$1300 AU).  

We present the results of ASTE 10 m telescope observations in CO(3--2), HCO$^+$, H$^{13}$CO$^+$(4--3), and 1.1 mm dust-continuum emissions and SMA observations in $^{12}$CO, $^{13}$CO, C$^{18}$O(2--1), and 1.3 mm dust-continuum emissions toward the low-mass Class-0/I protostar, B59\#11, which is thought to be one of the good targets for investigating the disk formation in the early protostellar evolution. First, we present the details of our ASTE observations and SMA data reductions in Section \ref{sec:observations}. In Section \ref{sec:results}, we show the results of ASTE and SMA observations and derive the physical properties of the outflow and the dense gas associated with B59\#11. In Section \ref{sec:discussions}, we discuss the possibility that B59\#11 has a rotationally supported disk and a rotating outflow. Finally, we summarize our main conclusions in Section \ref{sec:summary}.

\section{Observations}
\label{sec:observations}
\subsection{AzTEC/ASTE Observations}
\label{sec:obs:aztc}
We carried out 1.1 mm dust-continuum observations toward the B59 region with the AzTEC camera \citep{wil08} {mounted on} the ASTE 10 m telescope \citep{ezawa04,kono04} located at Pampa la Bola (altitude=4800 m), Chile. The observations were performed in the period October 17 to 31, 2008. The weather conditions during the period were good or moderate, and the typical atmospheric opacity at 225 GHz was in the range of 0.04-0.2. The AzTEC camera mounted on the ASTE telescope is a 144-element bolometric camera and provides us with an angular resolution of 28\arcsec\hspace{1ex}in full width at half maximum (FWHM)\citep{wil08}. The 1.1 mm continuum observations of B59 were performed as a part of the survey of nearby star forming regions \citep{kawa13}. The observations were performed in the raster scan mode toward the 35\arcmin$\times$ 35\arcmin\hspace{1ex}area centered on $(\alpha_{\rm J2000}, \delta_{\rm J2000}$)=(17$^{\rm h}$11$^{\rm m}$58\fs57,-27\arcdeg24\arcmin27\farcs86). Each field was observed several times with azimuth and elevation scans. The separation among scans was adopted to be 117\arcsec, which is a quarter of the AzTEC field of view (FoV; $\sim$7\arcmin.8 ). The scanning speed of the telescope was 250\arcsec\hspace{1ex}s$^{-1}$. {In total}, 28 individual maps {of} the entire field with an integration time of 9.4 minutes were taken, and those maps were averaged to produce {a} final map with a total integration time of 4.4 hours. The telescope pointing was checked every 2 hours by observing quasars, J1924-292 and J1733-130. The derived pointing offsets were linearly interpolated along the time sequence, and the interpolated pointing offset was applied to each target map. The pointing accuracy of the AzTEC map is estimated to be better than 2\arcsec. The flux scale was calibrated by observing the planet Uranus twice per night, and we measured the flux conversion factor (FCF) from optical loading value (in Watts) to the source flux (in Jy beam$^{-1}$) for each detector element. {A} principal component analysis (PCA : \citet{sco08}) cleaning method was applied to remove atmospheric noise. Details of the flux calibration are described by \citet{wil08} and \citet{sco08}. Since the PCA method does not have sensitivity to extended sources, we applied {an} iterative mapping method (FRUIT : \citet{liu10,shima11}) to recover the extended components. The noise level is $\sim$6 mJy beam$^{-1}$ in the central region and $\sim$7 mJy beam$^{-1}$ in the outer edge. The effective beam size of $\sim$36\arcsec\hspace{1ex}is estimated from Gaussian fitting of the point source in the map.
We also use the CLEANed image of PCA map to estimate parameters for point sources, in order to avoid the contamination due to the extended emissions seen in the FRUIT map. The PCA cleaning method {produced} a negative hole around the point-like source, due to the point-spread function (PSF) \citep[see ][]{tsuka11}. {A} CLEANed map was made by subtracting the measured PSF from emission via the CLEAN algorithm and by convolving the Gaussian beam with the FWHM of 35" to CLEAN components. Details are described in \cite{kawa13}. 

\subsection{ASTE $^{12}$CO(3--2), HCO$^+$(4--3), and H$^{13}$CO$^+$(4--3) Line Observations}
\label{sec:obs:aste}
We observed {the} $^{12}$CO($J=$3--2; 345.796 GHz), HCO$^+$($J=$4--3; 356.734 GHz), and H$^{13}$CO$^+$($J$=4--3; 346.998 GHz) transitions toward the B59 region during May in 2011 to January in 2012. The half-power beam width of the ASTE telescope is $\sim$22\arcsec\hspace{1ex}at the CO(3--2) frequency. The typical system noise temperature with the 345 GHz SIS heterodyne receiver was 300-600 K during our observations. The temperature scale was determined by the chopperwheel method \citep{kut81}, which provides us with the antenna temperature corrected for the atmosphere attenuation. As a back end, we used four sets of a 1024 channel auto-correlator, which {provided} frequency resolution of 125 kHz, corresponding to $\sim$0.1 km s$^{-1}$ at the HCO$^+$(4--3) and H$^{13}$CO$^+$(4--3) frequencies. The on-the fly (OTF) mapping technique was used to construct a map covering an area of 15\arcmin $\times$11\arcmin\hspace{1ex}(corresponding to 0.6$\times$0.4 pc) in CO(3--2) emission. In addition, the position-switching mode was used to construct two smaller {maps} of H$^{13}$CO$^+$ and HCO$^+$ emission, the first one with an area of 60\arcmin$\times$60\arcmin\hspace{1ex}(corresponding to 0.04$\times$0.04 pc), and the second one with an area of 80\arcmin$\times$80\arcmin\hspace{1ex}(corresponding to 0.05$\times$0.05 pc). In both maps, a grid separation of 20\arcsec\hspace{1ex}was used. The telescope pointing was checked every 2 hours by five-point scans of the point-like $^{12}$CO($J$ = 3--2) emission from Waql ($\alpha_{J2000.0}$,$\delta_{J2000.0}$)=(19$^{\rm h}$15$^{\rm m}$23\fs35 , -07\arcdeg02\arcmin50\farcs3), and IRAS 16594-4656 ($\alpha_{J2000.0}$,$\delta_{J2000.0}$)=(17$^{\rm h}$03$^{\rm m}$10\fs03 , -47\arcdeg00\arcmin27\farcs68). The pointing errors were measured to be from 1\arcsec\hspace{1ex}to 2\arcsec\hspace{1ex}during the observations. A main beam efficiency is 50\%. We subtracted linear baselines from the OTF spectra, then we convolved {the maps with} a spheroidal function and resampled them onto a 7\farcs5 grid. Since the telescope beam size is 22\arcsec, the effective FWHM resolution in the restored images is 27\arcsec. The scanning effect was minimized by combining scans along the R.A. and {Decl.} directions, using the PLAIT algorithm developed by \citet{emer88}. The typical rms noise level in the final image is 1.5 K in ${\rm T}_{\rm A}^*$ at a velocity resolution of 0.1 km s$^{-1}$.

\subsection{SMA Data Reduction}
\label{sec:obs:sma}
 We also processed the archival data from the SMA observations of [BHB2007]\#11 and constructed both the continuum and spectral line images. The 1.3 mm continuum emission was observed with SMA in the compact configuration (7 antennas). The minimum and maximum baselines are 7 k$\lambda$ and 50 k$\lambda$, respectively. $^{12}$CO($J$=2--1; 230.538 GHz), $^{13}$CO($J$=2--1; 220.399 GHz), and C$^{18}$O($J$=2--1; 219.560 GHz) emissions were observed simultaneously with the 1.3 mm continuum emission. The raw data were calibrated using the MIR package for IDL that was developed for reductions of SMA data based on the Owens Valley Radio Observatory MMA package \citep{scov93}. In the reduction, visibilities with clearly deviating phase and/or amplitudes were flagged. Observations of the calibrator NRAO530 and J1924-292 were interleaved with the target for complex gain calibration. The passband response was calibrated using a strong source 3C454.3. Absolute flux scale was determined by a bootstrap method with Callisto and should be accurate at the 10\% level, by comparison of the quasar fluxes with the SMA calibration database. The phase reference center toward the target is ($\alpha_{J2000.0}$,$\delta_{J2000.0}$)=(17$^{\rm h}$11$^{\rm m}$23\fs18, -27\arcdeg24\arcmin31\farcs5). After the calibrations, final CLEANed images were made using MIRIAD \citep{sau95} with natural uv weighting. The resulting synthesized beam size was 5\farcs0$\times$2\farcs8 (corresponding to 650$\times$360 AU) with a position angle of 31\arcdeg\hspace{1ex}for the dust-continuum map. The achieved rms noise levels were 12 mJy beam$^{-1}$ for the dust-continuum map and 200 mJy beam$^{-1}$ for $^{13}$CO, C$^{18}$O, and CO(2--1) images. The final images were uncorrected for the primary beam attenuation. The details of observations are summarized in Table. \ref{SMAobs}.

\section{Results}
\label{sec:results}
\subsection{Large Scale Structures of B59}
\subsubsection{AzTEC/ASTE 1.1 mm Dust-Continuum Emission}
\label{sec:res:aztc}
\begin{figure}
\epsscale{0.8}
\plotone{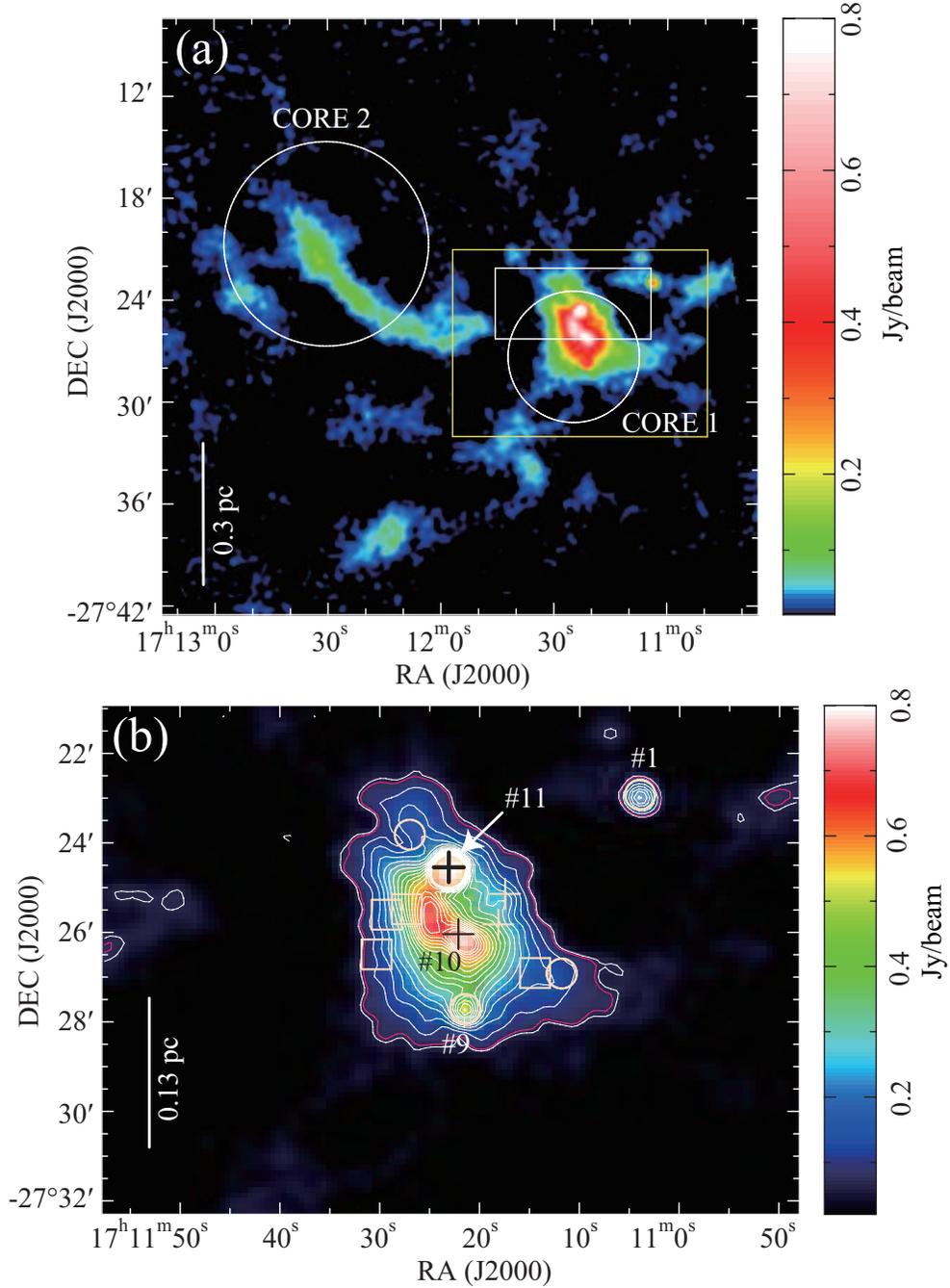}
\caption{AzTEC/ASTE 1.1 mm dust-continuum map. The white box shows the area of Figure \ref{COch}, and the yellow box shows the observing area of ASTE CO(3--2) emission (Figure \ref{COwing}). Contour levels start at 3$\sigma$ noise level with an interval of 6$\sigma$ (1$\sigma$=7 mJy beam$^{-1}$). The B59 mass is derived from emission within the pink contour ($\geq10\sigma$) (Section \ref{sec:res:aztc}). The crosses, circles, and boxes show positions of Class-I, flat, and Class-II YSOs, respectively \citep{for09}. The white circles in (a) show the locations and the extents of Core 1 and Core 2 identified by \citet{onishi99}}
\label{aztec}
\end{figure}

Figure \ref{aztec} shows the AzTEC/ASTE 1.1 mm FRUIT image obtained toward the north end of the Pipe Nebula. The image shows two dusty clumps which correspond to Core 1 (i.e., B59) and Core 2 detected in the low spatial resolution C$^{18}$O(1--0) map \citep{onishi99}. The clump associated with Core 2 has a filamentary structure $\sim0.6$ pc long. In addition, other filamentary structures were also detected in our map. These overall structures are in good agreement with the $A_V$ map produced by infrared observations, and the 1.2 mm dust-continuum map \citep{rom09,rom12}. The B59 clump consists of several dust condensations in 1.1 mm dust-continuum emission. Peak positions of four condensations coincide with positions of YSOs, [BHB2007]\#1, \#9, \#10, and \#11 identified in the infrared surveys \citep{bro07}. The dust-continuum emission associated with B59\#11 is the strongest in the B59 region, and has a peak intensity of 1.9 Jy beam$^{-1}$. On the northeast of B59\#11, a cavity-like structure $\sim$0.06 pc long was found to be elongated along the southwest-northeast direction. This feature coincides with {the one} found in the $A_V$ map \citep{rom09}. 

The mass of B59 (Core 1), $M_{\rm dust}$, was derived to be 24 $M_\sun$ from the 1.1 mm total flux obtained from the FRUIT image, $F_\nu$, using

  \begin{equation}
  M_{\rm dust}=\frac{F_\nu d^2}{\kappa_\nu B_\nu (T_d)}
  \label{eq:dust}
  \end{equation}
  
where we assume that all the 1.1 mm dust-continuum emission arises from the dust and is optically thin. Here, we adopt the mass opacity coefficient, $\kappa_\lambda=0.1(250\micron/\lambda)^{\beta}$ cm$^2$ g$^{-1}$ \citep{hil83} and $\beta$=2, which is a typical value in the interstellar medium \citep{kna73}. For the dust temperature, we adopted $T_{\rm d}$=10 K, which is derived from the observations in NH$_3$ lines \citep{rath08}. The estimated mass is in good agreement with that of 22.4 $M_\sun$ derived by \citet{rom09}  from the $A_V$ map. The mass of the dusty filament (corresponding to Core 2) is estimated to be 4.3 $M_\sun$ from the FRUIT map. 

\begin{figure}
\epsscale{1.0}
\plotone{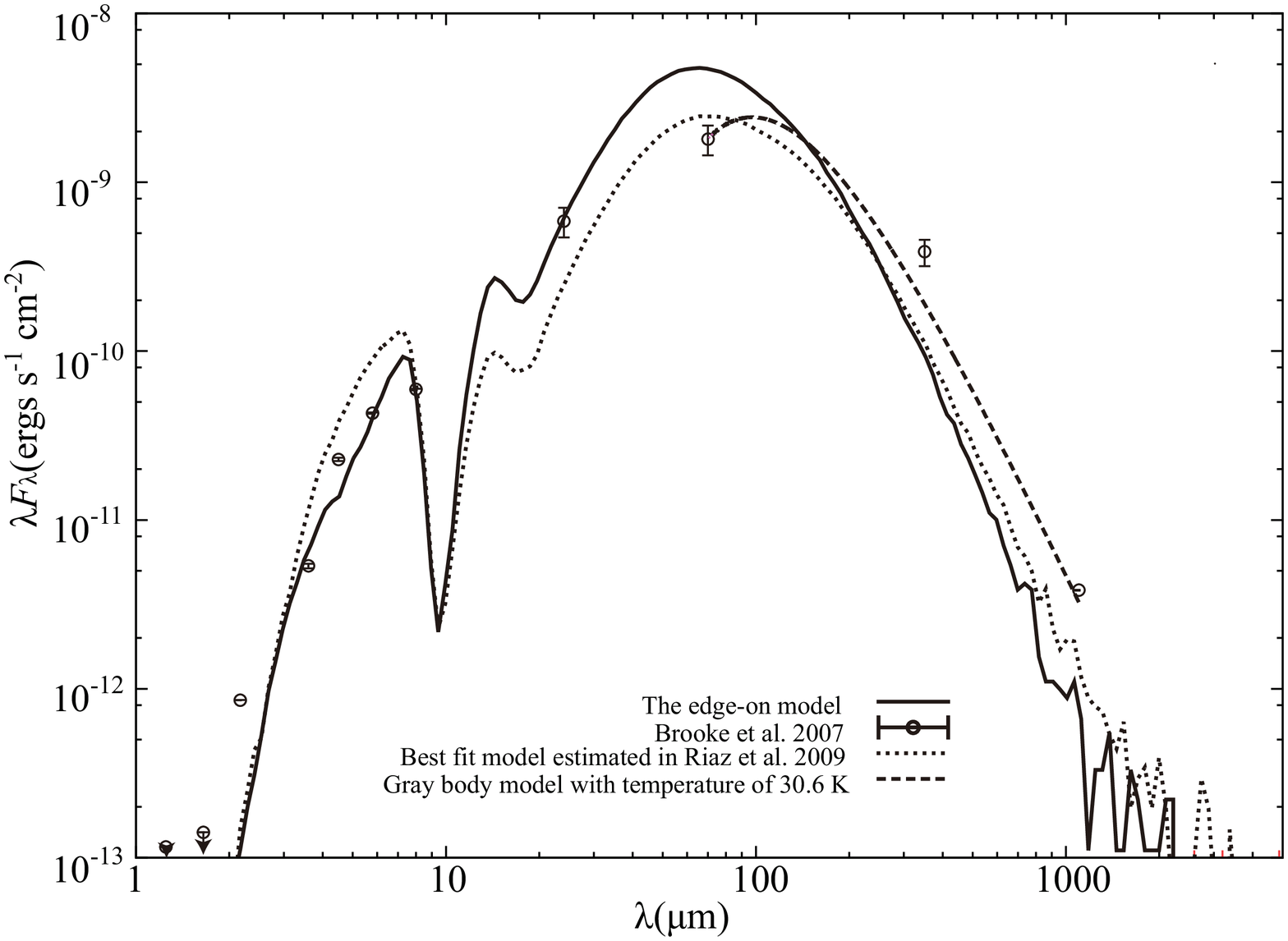}
\caption{Spectral energy distribution (SED) of B59\#11. The flux densities of infrared emissions are from \citet{bro07}, and that of the 350 \micron\hspace{1ex}emission is from the results of SHARC-II observations \citep{wu07}. We used the two-dimensional radiative transfer code by \citet{whit03} and the SED fitting tool by \citet{rob07}. The solid line shows the edge-on protostellar model and the dotted line shows the model fitted by \citet{riaz09}. The dashed line shows the graybody model with a temperature of 30.6 K and $\beta=1$. }
\label{sed}
\end{figure}

The mass of the dust condensation associated with B59\#11 is also estimated using Equation \ref{eq:dust}. Here, we used the CLEANed PCA map in order to avoid the contamination due to the extended emission seen in B59, since CLEANed PCA map is less sensitive to extended emissions than the FRUIT map (see \citet{sco08} for PCA and \citet{kawa13} for CLEANed PCA). The peak flux density of the dust condensation associated with B59\#11 is obtained to be 1.3 Jy beam$^{-1}$ from the CLEANed PCA map with an effective beam size of 36\arcsec. For optically thin emission, the spectral slope between two frequencies, $\alpha=\frac{\log{(F_{\nu1}/F_{\nu0})}}{\log{(\nu_1/\nu_0)}}$, will be related to the slope, $\beta$, of the dust opacity law, $\kappa\propto\nu^\beta$, as $\alpha\thickapprox\beta+2$ in the Rayleigh-Jeans limit \citep{beck91}. Using the flux estimated from the SHARC-II 350\hspace{1ex}\micron\hspace{1ex}map for a 40\arcsec\hspace{1ex}aperture of $F_{\rm 350\micron}$=45.2 Jy \citep{wu07} and $F_{\rm 1.1mm}$ in the AzTEC/ASTE observations, the spectral slope, $\alpha$, is estimated to be 3.0$^{+0.1}_{-0.2}$ and $\beta$ is estimated to be $\sim${1}. The dust temperature is estimated to be $\sim$31 K by a graybody fit with MIPS 70 \micron\hspace{1ex}\citep{bro07}, SHARC-II 350 \micron\hspace{1ex}\citep{wu07}, and AzTEC 1.1 mm emissions using $\beta$=1 (Figure \ref{sed}). The 70 \micron \hspace{1ex}flux possibly contains the emission from the central star, and this temperature gives an upper limit of the dust temperature. Using these values, the mass of the dust condensation associated with B59\#11 is estimated to be 0.09 $\pm$ $5.3\times10^{-4}$ $M_\sun$. The FWHM of the 1.1 mm dust condensation associated with B59\#11 is  37\arcsec$\times$33\arcsec (PA=-47\arcdeg), and the deconvolved size cannot be estimated, since the dust condensation is not resolved. \citet{rom12} {estimated} that the deconvolved size of the dust condensation associated with B59\#11 is 18\arcsec$\times$17\arcsec\hspace{1ex}from the 1.2 mm dust-continuum observations with a beam size of 11\arcsec. Using this value, the mean gas density, $n$, is estimated by assuming a spherically symmetric shape and using a geometric mean radius, as follows; 

\begin{equation}  
  n=\frac{M_{\rm dust}}{(4/3)\pi[\sqrt{(D_{\rm maj}/2)(D_{\rm min}/2)}]^3\mu_gm_{{\rm H}_2}}
  \label{eq:dense}
\end{equation}

$D_{\rm maj}$ and $D_{\rm min}$ are source sizes along major and minor axes, respectively. $\mu_g$ is mean atomic weight set to 1.36. The mean gas density is estimated to be $n=1.9\times10^6$ cm$^{-3}$.

 \subsubsection{ASTE $^{12}$CO(3--2) Emission}
 \label{sec:res:asteco}
 Our ASTE CO(3--2) data show that the CO(3--2) line profiles around B59\#11 have high velocity wings that probably originate from the molecular outflow ejected from B59\#11 (Figure \ref{COpro}). Here, we identify {a} molecular outflow from B59\#11 using the ASTE CO(3--2) data, and derive the physical properties of the outflow (Table \ref{COpar}). 
 
\begin{figure} 
\epsscale{1.0}
\plotone{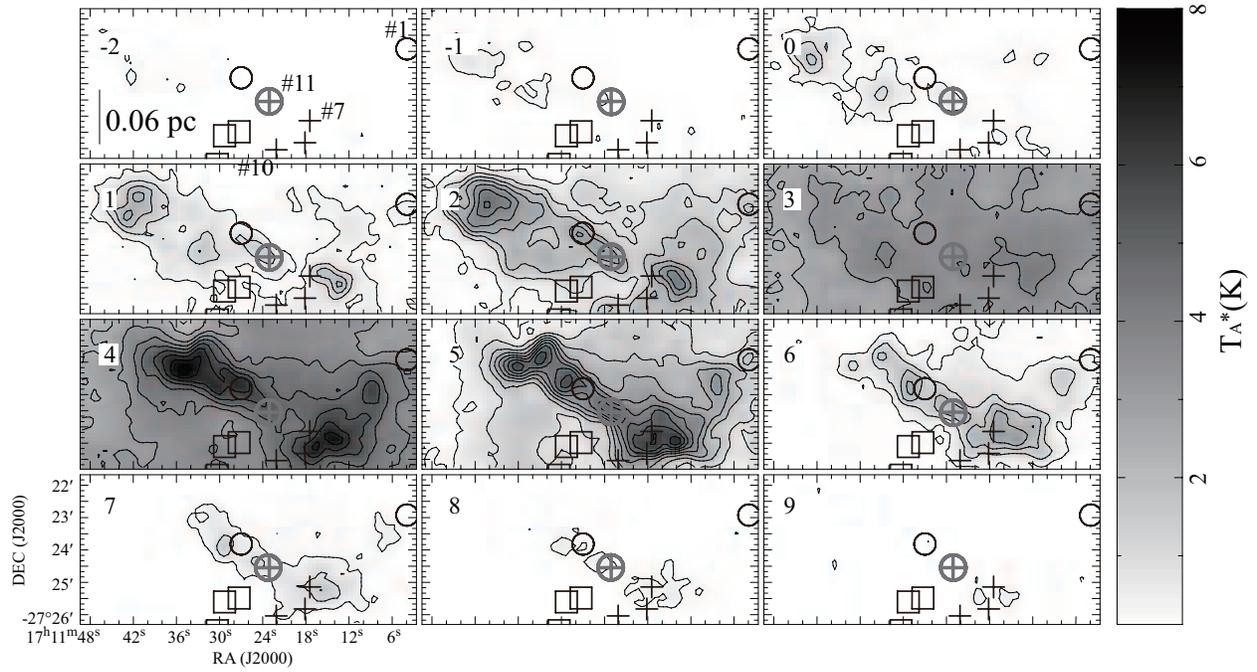}
\caption{Velocity channel maps of ASTE CO(3--2) emission with the velocity range of -2-9 km s$^{-1}$. Contour levels start at 3$\sigma$ noise level with an interval of 3$\sigma$ (1$\sigma$=0.2 K($T_{\rm A}^*$)). The symbols are the same as those in Figure. \ref{aztec}. {The gray circles around B59\#11 is the SMA FoV. }}
\label{COch}
\end{figure}

From the inspection of the CO(3--2) velocity channel maps (Figure \ref{COch}) and the profile map (Figure \ref{COpro}), we consider that the cloud component has {a} velocity range of $V_{\rm LSR}=1.5-5.5$ km s$^{-1}$. The components with velocities smaller than $V_{\rm LSR}=1.5$ km s$^{-1}$ and larger than 5.5 km s$^{-1}$ are considered to correspond to the blueshifted and redshifted molecular outflow components, respectively.
The main characteristics of the CO(3--2) data are summarized as follows;
\begin{enumerate}[a{)}]
\item Both the blueshifted and redshifted emissions are seen on the northeast of B59\#11 with lengths of $\sim$0.2 pc and $\sim$0.1 pc, respectively.
\item High velocity emission {is} mostly centered on B59\#11.
\item The CO(3--2) profiles have high-velocity wings that probably originate from the outflows from the YSOs, [BHB2007]\#1, \#9, and \#10. The outflow lobes of these YSOs can be recognized in Figure \ref{COch}.
\end{enumerate} 
The high velocity wings associated with B59\#11 noted above a) are roughly aligned in the same direction as the extended nebulosity seen in IRAC 3.6 \micron\hspace{1ex}and 4.5 \micron\hspace{1ex}images \citep{bro07}, indicating that these high velocity components are the outflow wings. The cavity-like structure on the northeast of B59\#11 seen in the AzTEC/ASTE 1.1 mm dust-continuum image coincides with the outflow; especially redshifted emission on the northeast of B59\#11 is possibly well fit to the cavity. The characteristics, a) and b) {above,} indicate that the outflow associated 
with B59\#11 is nearly along the plane of the sky direction and dominates the high velocity emission in the B59 region. These overall features are in good agreement with the CO(3--2) maps in \citet{dua12}.

Figure \ref{COpro} (b) shows a profile map of the 80\arcsec$\times$80\arcsec\hspace{1ex}area around B59\#11 with a grid spacing of 20\arcsec. Strong redshifted wing (and weaker blueshifted emission) exists on the southwest of B59\#11, while both the blueshifted and redshifted emissions are seen on the northeast. 

 \hspace{1ex} We derived physical properties of the outflow using {an} excitation temperature of 25 K and an outflow inclination angle of 75\arcdeg\hspace{1ex}following \citet{dua12}. We assumed the local thermal equilibrium (LTE) condition and {used} the following equation;  
 \begin{eqnarray}
M_j&=&\mu_gm_{{\rm H}_2}X^{-1}_{\rm CO}\Omega d^2 \sum_i N_{{\rm CO},i,j}
\\ &=&4.65\times10^{-7}\left (\frac{X_{\rm CO}}{1.0\times10^{-4}}\right )^{-1}\left (\frac{d}{130{\rm pc}} \right )^2\left (\frac{\theta}{22\arcsec}\right )^2\left (\frac{\eta}{0.5} \right )^{-1} \nonumber
\\&\hspace{2ex}& \sum_i\left (\frac{T_{{\rm A},i,j}^*}{\rm K\cdot km\hspace{1ex}s^{-1}}\right )\left (\frac{T_{\rm ex}}{25\rm K}\right )\exp{[33.2/T_{\rm ex}]} M_{\sun}
\label{eq:co32}
\end{eqnarray}
 The mass, momentum, and kinetic energy of the outflow are summarized in Table. \ref{COpar}. 
 
 \subsection{The Dense Gas Distribution in B59\#11}
 
\subsubsection{ASTE HCO$^+$ and H$^{13}$CO$^+$(4--3) observations}
\label{sec:res:astehco}
Here, we present the dense gas distribution associated with B59\#11. Figure \ref{HCO+pro} shows a 5$\times$5 points profile map in ASTE HCO$^+$(4--3) emission (black lines) taken toward B59\#11 with a grid spacing of 20\arcsec, together with a 3$\times$3 points profile map in H$^{13}$CO$^+$(4--3) emission (red lines) with the same grid spacing. The HCO$^+$(4--3) emission extends over the 80\arcsec$\times$80\arcsec\hspace{1ex}area. The strongest emission exists on the southwest of B59\#11, (-40,-20), and has a peak temperature of $\sim2.5\pm0.12$ K($T_{\rm A}^*$). The dotted line in Figure \ref{HCO+rat} (b) shows the residual spectrum after subtracting an average of the surrounding four spectra from that at the center. A high velocity component ($V_{\rm LSR}$=-1.5-2.5 km s$^{-1}$, 5.0$\sim$8.0 km s$^{-1}$) exists in both the central and residual spectra. After comparing both spectra, we found that the high velocity component is mostly spatially unresolved with the ASTE 22\arcsec\hspace{1ex}beam (corresponding to $\sim$3000 AU).  The velocity width of the high velocity component in HCO$^+$ emission is almost the same as that of the dense gas rotation, $V_{\rm LSR}$=-0.5-8.2 km s$^{-1}$ in SMA $^{13}$CO ($V_{\rm LSR}$=0.0-7.7 km s$^{-1}$) emission (see Section \ref{sec:res:smacoi}), and it implies that the high velocity component corresponds to the rotating dense gas. 

The velocity shift of 0.42 km s$^{-1}$ is detected along the direction from the northwest (20\arcsec,-20\arcsec) to the southeast (-20\arcsec,20\arcsec) of B59\#11 in the HCO$^+$ line (Figure \ref{HCO+rat}), and this direction coincides with that of the velocity shift {associated with} dense gas rotation {as inferred from the} SMA $^{13}$CO and C$^{18}$O emissions (Section \ref{sec:res:smacoi}). This means that the velocity shift in HCO$^+$ emission {possibly arises} from the large-scale envelope rotation. The specific angular momentum of the outer envelope is estimated to be 3.7$\times10^{-3}$ km s$^{-1}$ pc if the shift is due to the rotation.

The H$^{13}$CO$^+$(4--3) intensity is strongest at the center ($T_{\rm A}^*\sim$0.11 K) and no clear signature of self-absorption is found. The peak velocity and the FWHM velocity width of the central spectrum are obtained to be 3.6 km s$^{-1}$ and as wide as 1.7 km s$^{-1}$. The peak velocity of 3.6 km s$^{-1}$ is adopted {as} the systemic velocity of B59\#11 in this paper. The H$^{13}$CO$^+$(4--3) spectra are very narrow at other positions; e.g., positions at (0\arcsec,20\arcsec) and (20\arcsec,-20\arcsec).

From the comparison of the HCO$^+$ and H$^{13}$CO$^+$ line profiles, the emissions in H$^{13}$CO$^+$ correspond to the dips in HCO$^+$ in velocity; a dip is located at 3-4 km s$^{-1}$ and emission is located at 2-5 km s$^{-1}$ at the center. It indicates that the dips are formed by self-absorption. On the other hand, the HCO$^+$ intensities at the blueshifted parts are stronger than the redshifted parts on the southwest of B59\#11, for example, (-20\arcsec,0\arcsec) and (-20\arcsec,-20\arcsec) in Figure \ref{HCO+pro}. These "blue-skewed" profiles are usually considered to indicate dynamical infall \citep[e.g.,][]{zhou93}. It is noted that a red-skewed profile is obtained at the central part of B59\#11, and a simple infall model cannot account for the overall observational results.

Masses of the high velocity components in HCO$^+$ emission are estimated using the following equation on the assumption of LTE condition. 
 \begin{eqnarray}
M_{\rm HCO^+}&=&\mu_gm_{{\rm H}_2}X^{-1}_{\rm HCO+}\Omega d^2 \int N_{{\rm HCO^+}}dv
\\ &=&1.03\times10^{-2}\left (\frac{X_{\rm HCO^+}}{1.0\times10^{-10}}\right )^{-1}\left (\frac{d}{130{\rm pc}} \right )^2\left (\frac{\theta}{22\arcsec}\right )^2\left (\frac{\eta}{0.5} \right )^{-1} \nonumber
\\&\hspace{2ex}& \int\left (\frac{T_{{\rm A}}^*}{\rm K\cdot km\hspace{1ex}s^{-1}}\right )dv\left (\frac{T_{\rm ex}}{30\rm K}\right )\exp{[42.8/T_{\rm ex}]} M_{\sun}
\label{eq:hco+}
\end{eqnarray}
where we assume that HCO$^+$ emission is optically thin in the high velocity ranges.  We used $X[{\rm HCO^+}]=1.0\times10^{-10}$ \citep{raw04}, and $T_{\rm ex}$=30 K, which we adopt for the envelope/disk in this paper (Section \ref{sec:res:aztc}) since its high velocity component is considered to trace the rotating dense gas. The masses of the blueshifted ($V_{\rm LSR} = -1.5-2$ km s$^{-1}$) and redshifted ($V_{\rm LSR}=5-8$ km s$^{-1}$) components are estimated to be 1.2$\times10^{-2}$ and 1.8$\times10^{-2}$ $M_\sun$, respectively. The total mass of the high velocity components, 0.03 $M_\sun$, is comparable to the masses obtained from SMA $^{13}$CO and C$^{18}$O data. {This indicates} that the high velocity components trace the dense gas rotation in terms of the mass comparison as well as the comparison of the velocity width described above. 

The mass of dense region traced by H$^{13}$CO$^+$(4--3) is estimated on the assumption of LTE condition.
 \begin{eqnarray}
M_{\rm H^{13}CO^+}&=&\mu_gm_{{\rm H}_2}X^{-1}_{\rm H^{13}CO+}\Omega d^2 \int N_{{\rm H^{13}CO^+}}dv
\\ &=& 7.8\times10^{-1}\left (\frac{X_{\rm H^{13}CO^+}}{1.4\times10^{-12}}\right )^{-1}\left (\frac{d}{130{\rm pc}} \right )^2\left (\frac{\theta}{22\arcsec}\right )^2\left (\frac{\eta}{0.5} \right )^{-1} \nonumber
\\&\hspace{2ex}& \int\left (\frac{T_{{\rm A}}^*}{\rm K\cdot km\hspace{1ex}s^{-1}}\right )dv\left (\frac{T_{\rm ex}}{30\rm K}\right )\exp{[41.6/T_{\rm ex}]} M_{\sun}
\label{eq:h13co+}
\end{eqnarray}
We adopt $X[{\rm H^{13}CO^+}]=1.4\times10^{-12}$ estimated with a C to $^{13}$C abundance ratio of $\sim71$ \citep{wil94}. H$^{13}$CO$^+$ emission is assumed to be optically thin, and we obtained a mass of 0.59 $M_\sun$. These results are summarized in Table. \ref{HCO+par}. 
 
 \subsubsection{SMA $^{13}$CO(2--1) and C$^{18}$O(2--1) Emission}
\label{sec:res:smacoi}
\begin{figure} 
\epsscale{1.0}
\plotone{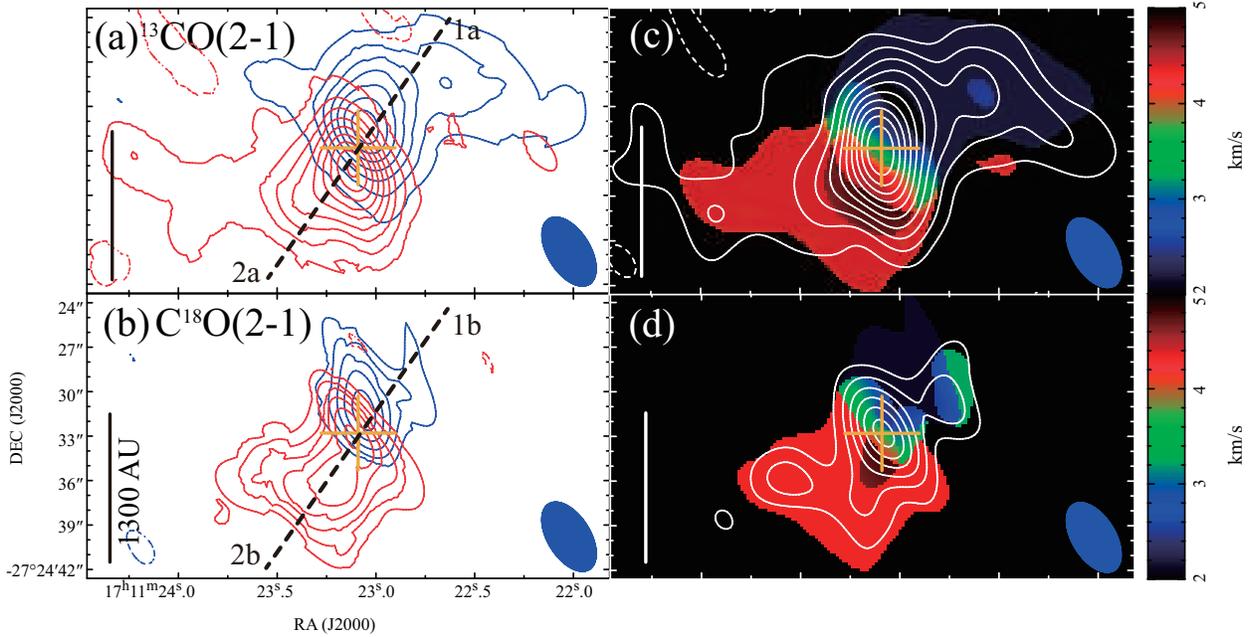}
\caption{The maps of (a) (c) SMA $^{13}$CO(2--1) and (b) (d) C$^{18}$O(2--1) emissions. The blue and red contours represent blueshifted and redshifted intensities integrated between (a) $V_{\rm LSR}$=-0.5 to 1.2 km s$^{-1}$ and 3.9 to 8.2 km s$^{-1}$, and (b) 0.5 to 2.6 km s$^{-1}$ and 3.8 to 8.2 km s$^{-1}$, respectively. The white contours superposed on the mean velocity maps in (c) and (d) indicate the emissions integrated over intervals from $V_{\rm LSR}$=-0.5 to 8.2 km s$^{-1}$, and from 0.5 to 8.2 km s$^{-1}$. Contour levels start at 3$\sigma$ noise level with intervals of (a) 3$\sigma$ (1$\sigma$=0.5 Jy beam$^{-1}$ km s$^{-1}$), (b) 1.5$\sigma$ (1$\sigma$=0.3 Jy beam$^{-1}\cdot$km s$^{-1}$, (c) 3$\sigma$ (1$\sigma$=0.6 Jy beam$^{-1}\cdot$km s$^{-1}$, and (d) 1.5$\sigma$ (1$\sigma$=0.6 Jy beam$^{-1}\cdot$km s$^{-1}$). The blue ellipses at the bottom-right corners show the beam size, and the crosses indicate the position of B59\#11 \citep{for09}. The dashed lines in (a) and (b) indicate the cuts along which the P-V diagrams in Figure \ref{pv} (a) and (b) were produced.}
\label{smadense}
\end{figure}

In this section, we show the results of SMA $^{13}$CO and C$^{18}$O(2--1) observations. These emissions are expected to trace the smaller scale of the dense gas than that obtained by ASTE HCO$^+$ and H$^{13}$CO$^+$ observations. From the total integrated intensity maps (Figure \ref{smadense} (c) and (d)), it is shown that a compact gas condensation in $^{13}$CO and C$^{18}$O emissions is clearly associated with B59\#11 and elongated along northwest-southeast direction, which is perpendicular to the molecular outflow axis identified with ASTE CO(3--2) observations (Section \ref{sec:res:asteco}). The size of the C$^{18}$O condensation is about half the size of the $^{13}$CO condensation, and {its} extent {are} measured to be 17\arcsec$\times$11\arcsec (corresponding to $\sim$2000$\times$1400 AU and an aspect ratio of $\sim$1.5) and 30\arcsec$\times$17\arcsec (corresponding to $\sim$4000$\times$2000 AU and an aspect ratio of $\sim$1.8), respectively. 

A velocity gradient is evident in both the $^{13}$CO and the C$^{18}$O maps (Figure \ref{smadense}) along the major axis of the dense gas condensation. The northwestern side of the condensation is blueshifted and the southeastern side is resdhifted. Figure \ref{pv} (a) and (b) show the Position-Velocity (P-V) diagrams which are cut along the major axis of the condensation. The velocity gradient {appears to have a} power-law profile, indicating that the velocity gradient is arisen from the dense gas rotation. The specific angular momentum is estimated to be 2.1$\pm1.2\times10^{-3}$ km s$^{-1}$ pc using the results of $^{13}$CO(2--1) assuming the inclination angle of 75\arcdeg\hspace{1ex}\citep{dua12}. 

The optical depth of $^{13}$CO(2--1) emission is calculated assuming that the abundance ratio between $^{13}$CO and C$^{18}$O is $\sim$6 \citep{fre87} and use the following equation,  
\begin{equation}
\frac{T^*_{\rm R}({\rm ^{13}CO})_v}{T^*_{\rm R}({\rm C^{18}O})_v}=\frac{1-\exp[-\tau_v({\rm ^{13}CO})]}{1-\exp[-\tau_v({\rm ^{13}CO})/6]}
\label{eq:tau}
\end{equation}
where $T_{\rm R}^*$ is the radiation temperature at a given velocity, $v$. The $^{13}$CO emission is estimated to be optically thick for only two velocity channels that are closed to the systemic velocity, $V_{\rm LSR}=3.3-4.4$ km s$^{-1}$. Noted that emissions in the channels severely suffer from resolved-out effect in both lines{,} and {thus our} estimate of the optical depth includes a large uncertainty due to {such} effect.
On the assumption of LTE condition, we also obtain the mass of dense gas traced by $^{13}$CO(2--1) emission as follows;
\begin{eqnarray}
M_{\rm ^{13}CO}&=&\sum_j M_j=5.11\times10^{-5}\left (\frac{X_{^{13}{\rm CO}}}{1.7\times10^{-6}}\right )^{-1}\left (\frac{d}{130{\rm pc}} \right )^2 \nonumber
\\ &\hspace{2ex}&\sum_i\left (\frac{S_{\nu,i,j}}{\rm Jy\cdot km\hspace{1ex}s^{-1}}\right )\left (\frac{T_{\rm ex}}{30\rm K}\right )\exp{[15.9/T_{\rm ex}]}\frac{\tau_{\rm ^{13}CO}}{1-e^{-\tau_{\rm ^{13}CO}}}\hspace{1ex}M_{\sun}
\label{eq:13co}
\end{eqnarray}
We used $T_{\rm ex}=30$ K and $X[\rm ^{13}CO]=1.7\times10^{-6}$ \citep{fre87}. From a total integrated flux of 30.4 Jy$\cdot$km s$^{-1}$, the mass is estimated to be 3.7$\times10^{-2}$ $M_\sun$. 
On the assumption of LTE condition and optically thin emission \citep{fre82}, we compute the mass of dense gas traced by C$^{18}$O(2--1) emission as follows;
\begin{eqnarray}
M_{\rm C^{18}O}&=&\sum_j M_j=3.03\times10^{-4}\left (\frac{X_{\rm C^{18}O}}{3.0\times10^{-7}}\right )^{-1}\left (\frac{d}{130{\rm pc}} \right )^2 \nonumber
\\ &\hspace{2ex}&\sum_i\left (\frac{S_{\nu,i,j}}{\rm Jy\cdot km\hspace{1ex}s^{-1}}\right ) \left (\frac{T_{\rm ex}}{30\rm K}\right )\exp{[15.6/T_{\rm ex}]}\hspace{1ex}M_{\sun}
\label{eq:c18o}
\end{eqnarray}
We adopt $T_{\rm ex}$=30 K and $X[\rm C^{18}O]=3.0\times10^{-7}$ \citep{fre87}. Using a total integrated flux of 8.9 Jy$\cdot$km s$^{-1}$, the mass is derived to be $3.3\times10^{-2}$ $M_\sun$.
 
\subsection{SMA 1.3 mm Dust-Continuum Emission}
\label{sec:res:smacon}
\begin{figure} 
\epsscale{1.0}
\plotone{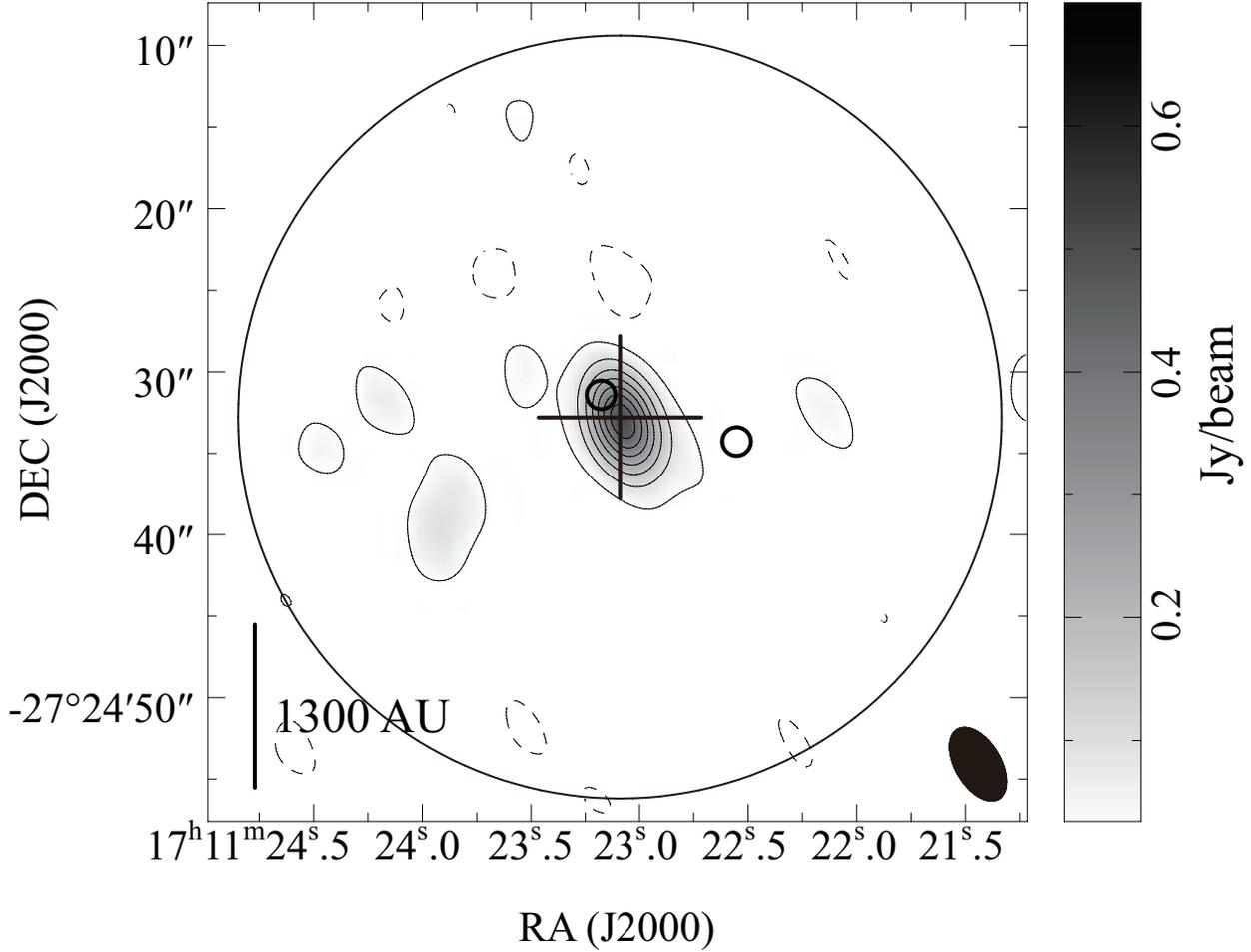}
\caption{SMA 1.3 mm dust-continuum map. Contour levels are start at 3$\sigma$ noise level with intervals of 6$\sigma$ (1$\sigma$=0.012 Jy beam$^{-1}$). The small circles show the positions of B59\#11 and B59\#11SW which are identified with the infrared observations \citep{riaz09}. The big circle shows the SMA FoV of 54\arcsec. The other symbols are the same as those in Figure \ref{smadense}.}
\label{C11}
\end{figure}

Figure \ref{C11} shows the distribution of 1.3 mm dust-continuum emission obtained by SMA. A compact and strong dust condensation associated with B59\#11, centered at ($\alpha_{J2000.0}$,$\delta_{J2000.0}$)=(17$^{\rm h}$11$^{\rm m}$23\fs08 , -27\arcdeg24\arcmin33\farcs1), has been detected. The FWHM size of the dust condensation is measured to be 5\farcs5$\times$3\farcs5 (PA=22\arcdeg), and its deconvolved size is estimated to be 2\farcs7$\times$1\farcs4 ($\sim$ 350$\times$ 180 AU) (PA=-15\arcdeg). The condensation is oriented along the same direction with the elongation of the dense gas distributions in SMA $^{13}$CO and C$^{18}$O emissions (see Section 3.6 and Section 4.1). The peak intensity and total integrated flux density are 0.49$\pm 0.012$ Jy beam$^{-1}$ and 0.67 Jy, respectively. The mass and mean gas density {are} estimated to be $\sim$7.3$\times10^{-2}$ M$_\sun$ and 1.1$\times$10$^9$ cm$^{-3}$, using Equation (1) on the same assumptions as those in the estimate for the AzTEC/ASTE 1.1 mm data (see Section 3.1). The separation between the position of B59\#11 defined by \citet{for09} and the SMA peak position is $\sim$0\farcs3, {i.e.} smaller than the position accuracy of the infrared images. At the position of B59\#11SW, no dust condensation has been detected and the disk or envelope mass associated with B59\#11SW is estimated to be lower than 4.9$\times10^{-4}$ $M_\sun$ (3$\sigma$ noise level) with $\beta=2$. {This} suggests that B59\#11SW is not embedded in the massive envelope and {may be} considered to be a more evolved source.

\subsection{SMA $^{12}$CO(2--1) Emission \\--- The velocity Gradient in the Molecular Outflow ---}
\label{sec:res:smaco}
We have detected an interesting internal structure in the outflow associated with B59\#11 as shown below.

Figure \ref{smaout} shows blueshifted ($V_{\rm LSR}=$-4.9-1.4 km s$^{-1}$) and redshifted ($V_{\rm LSR}=$4.6-12 km s$^{-1}$) components detected by SMA CO(2--1) observations. We {identified} three blueshifted and redshifted lobes in Figure \ref{smaout}; {one} blueshifted and {one} redshifted lobes are {projected towards} the northeastern side of B59\#11, while {the third, redshifted} lobe is {projected toward} the southwestern side. The lengths of {the} lobes are $\sim$2400 AU, 3100 AU, and 2400 AU, respectively. In both maps of ASTE CO(3--2) and SMA CO(2--1), the blueshifted and redshifted components exist on the northeastern side of B59\#11, while only a redshifted component exists on the southwestern side. This {shows} that the high velocity components observed in the SMA CO map trace the molecular outflow ejected from B59\#11. The mean velocity map of SMA CO emission (Figure \ref{smaout} (b)) shows the velocity gradient in the outflow lobe located on the northeast; the northern side of the outflow lobe is blueshifted and the southern side is redshifted. The direction of the velocity gradient in the outflow is the same as that of the dense gas rotation shown in Section \ref{sec:res:smacoi}. 

The mass, momentum, and energy are estimated from the following equation;
\begin{eqnarray}
M_{\rm ^{12}CO}&=&\sum_j M_j=6.09\times10^{-7}\left (\frac{X_{^{12}{\rm CO}}}{1.0\times10^{-4}}\right )^{-1}\left (\frac{d}{130{\rm pc}} \right )^2 \nonumber
\\ &\hspace{2ex}& \sum_i\left (\frac{S_{\nu,i,j}}{\rm Jy \cdot km\hspace{1ex}s^{-1}}\right ) \left (\frac{T_{\rm ex}}{25\rm K}\right )\exp{[16.6/T_{\rm ex}]}\hspace{1ex}M_{\sun}
\label{eq:co21}
\end{eqnarray}
We apply the same parameters as those of Equation \ref{eq:co32} in Section \ref{sec:res:asteco} and assume that CO(2--1) emission is optically thin. The masses for the blueshifted and redshifted outflows are derived to be $2.8\times10^{-4}$ and $4.4\times10^{-4}$ $M_\sun$, respectively. It is noted that CO lines often become optically thick even toward outflow wing components, and {thus, the} estimates of the outflow mass, momentum, or energy {are probably lower limits}. Moreover, it is shown that {roughly 90 \% of the emission is} resolved-out in SMA data, by comparison with the flux density estimated from ASTE CO(3--2) emission. The energies and momentums are also summarized in Table. \ref{COpar} together with the masses.

\section{Discussion}
\label{sec:discussions}
\subsection{Kinematics and physical properties of the rotating envelope and disk} 
\label{sec:dis:disk}

\subsubsection{Kinematical evidence for the formation of the Keplerian disk}
\label{sec:dis:linedisk}
As described in Section \ref{sec:res:smacoi}, $^{13}$CO and C$^{18}$O(2--1) emissions are oriented perpendicular to the outflow direction and also show the velocity gradient along its major axis, probably tracing a rotating flattened envelope or a disk of dense gas. The specific angular momentum of this system is as large as 2.1$\times10^{-3}$ km s$^{-1}$ pc (as described in Section \ref{sec:res:smacoi}). It is, therefore, expected that the centrifugal radius ($=j_n^2/GM_\ast$), which is {the} radius at the outer edge of the rotationally supported disk, is also as large as 200-1000 AU for e.g., $M_\ast=0.2-1$ $M_\sun$ and such a Keplerian disk can be identified. Here, we discuss whether the velocity gradient traces the rotation of the Keplerian disk or not in terms of the radial dependence of the rotation velocity. 

\begin{figure}
\epsscale{1.0}
\plotone{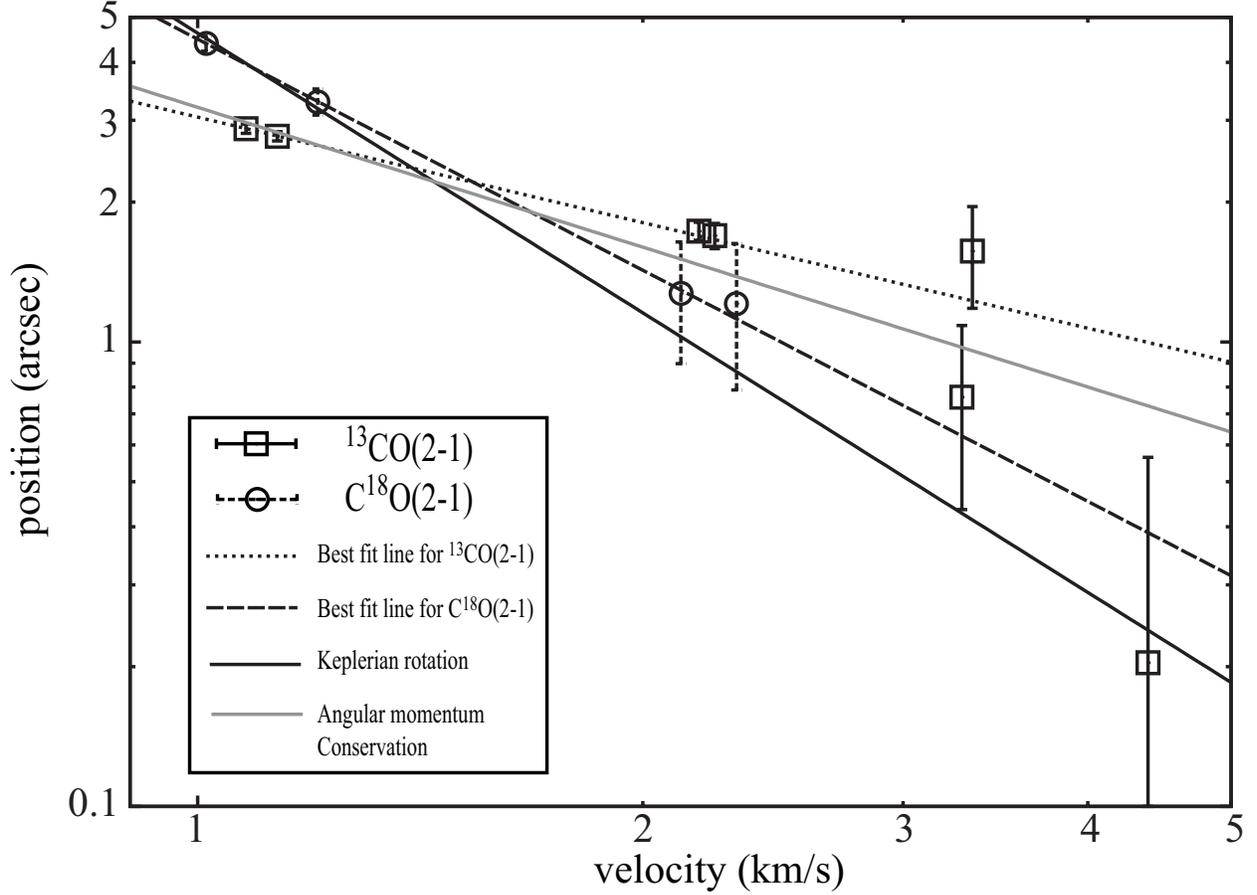}
\caption{Logarithmic plot of the measured radius as a function of velocities. The boxes and circles show separations of the positions between 1.3 mm dust-continuum peak and $^{13}$CO(2--1) and C$^{18}$O(2--1) peaks in each velocity channel. The dotted line shows the best fit line for $^{13}$CO(2--1) emission ($v\propto r^{-1.3}$) and the long-dashed line shows the best fit line for C$^{18}$O(2--1) emission ($v\propto r^{-0.61}$). The black solid line shows the Keplerian rotation with a central stellar mass of  0.73 $M_\sun$ fitted for C$^{18}$O(2--1) emission, and the gray solid line shows the rotation conserving its angular momentum of 2.1$\times10^{-3}$ km s$^{-1}$ pc.}
\label{velpos}
\end{figure}

Investigating the radial dependences of the rotational velocities allows us to discriminate two possible kinematics; the velocities {should be} proportional to $r^{-1}$ and $r^{-1/2}$ for a rotation conserving its angular momentum and a Keplerian rotation, respectively. We plotted radii as a function of velocities in $^{13}$CO and C$^{18}$O emissions as shown in Figure \ref{velpos}. The radii are estimated by measuring the peak positions of emissions and deriving the separations from the 1.3 mm dust-continuum peak. Via $\chi^2$ minimizations, the power-law indexes, $\alpha$ (i.e., $v=r^\alpha+v_{\rm sys}$) in $^{13}$CO and C$^{18}$O emissions are estimated to be -1.3 and -0.61, respectively (details are shown in Table \ref{fit}). These results suggest that the outer and lower density regions traced by the $^{13}$CO  are in better agreement with the presence of a rotating-flattened envelope, while the inner, more dense regions traced by C$^{18}$O emission (or, at least, some parts of this region) are in better agreement with the presence of a rotationally supported disk. 

The existence of a disk traced by C$^{18}$O can be also inferred from comparison between a dynamical mass and a stellar mass obtained from the SED analysis. To derive the dynamical mass of B59\#11, we fit Keplerian rotation (i.e., $v(r)=\sin{i}\sqrt{\frac{GM_\ast}{r}}$) to the C$^{18}$O data (Figure \ref{velpos}). The central stellar mass is estimated to be 0.73$^{+0.53}_{-0.39}$ $M_\sun$ assuming the inclination angle, $i$, of 75\arcdeg. This is an upper limit {to} the central mass {value}, since this inclination angle is determined by the limitation of the outflow opening angle {which} gives a lower limit. The central stellar mass estimated from this fitting is not consistent with the 0.28 $M_\sun$ obtained from the SED model fitting conducted by \cite{riaz09} using the radiative transfer model of \citet{whit03} for $i$=53\arcdeg-59\arcdeg. Our observations, however, reveal that the outflow associated with B59\#11 is ejected mostly in the plane of the sky (Section \ref{sec:res:asteco}). We also run the SED model fitting program developed by \citet{rob07} assuming $i>$75\arcdeg, and obtained the central stellar mass of $0.81$ $M_\sun$ as the best fit result. This model gives {a} better fit to data at the shorter wavelengths, $\lambda<$70 \micron, where emission comes {mostly} from the central protostar or the central region, compared with the model of \citet{riaz09} (Figure \ref{sed}). {The fit is}, however, worse at the longer wavelengths, especially at 70 \micron\hspace{1ex}and this might be caused by the contamination of B59\#11SW due to the larger observing beams (8\arcsec-36\arcsec). The estimated stellar mass from this SED analysis is well agreed with the dynamical mass obtained from SMA C$^{18}$O data, and this suggests a rotationally supported Keplerian disk traced by C$^{18}$O emission.   

\subsubsection{Evidence for disk formation from 1.3 mm image}
\label{sec:dis:contdisk}
 
Next, we also discuss whether the SMA 1.3 mm dust-continuum emission traces the disk/envelope system or exclusively the envelope. Here, we follow \cite{jor09}. They have pointed out that the SMA flux density (in 1.1 mm continuum emission) is only several \% of that of single dish observations (in 850 \micron\hspace{1ex}continuum emission) if there is no disk contribution about a protostar located at a distance of 125 pc. Here, we convert the distance and wavelengths in our observations to match with their method. Using the SMA data with the baselines $\geq 52$ k$\lambda$, the SMA to AzTEC flux density ratio is estimated to be 16$\pm$2\% where the flux densities are converted using $\beta=1$ (Section \ref{sec:res:aztc}). This value is large enough to be compared with 8\%, which is the largest value of the envelope contribution toward the various envelope models, suggesting the existence of the disk. We also note that the deconvolved size of the 1.3 mm dust condensation, 2.8\arcsec($\sim$350 AU), is consistent with an expected disk size of approximately 300 AU, which is obtained from the estimation of the centrifugal radius using the central stellar mass of 0.73 $M_\sun$ (Section \ref{sec:dis:linedisk}). The position angle is also consistent with the direction of the velocity gradient, would be another sign that the 1.3 mm dust-continuum emission traces the disk.

The SMA emission could be separated to contributions from the disk and the envelope, and individual masses are estimated using the following equations \citep{jor09}.
\begin{eqnarray}
S_{\rm 50k\lambda}=S_{\rm disk}+c\cdot S_{\rm env} \label{eq:dust1}\\
S_{15\arcsec}=S_{\rm disk}+S_{\rm env} \label{eq:dust2}
\end{eqnarray}
$c$ is a fraction of an envelope contribution to emission obtained by interferometric observations. On the assumption of $c$=0.04 \citep{jor09} and the same parameters as those in Section \ref{sec:res:aztc} and Section \ref{sec:res:smacon}, the envelope and disk masses are estimated to be 8.2$\times10^{-2}$ and 1.5$\times10^{-2}$ $M_\sun$, respectively.

From the above discussion, we conclude that, in source B59\#11, a rotationally supported disk {has} formed {and it is} in an early evolutionary phase. 

\subsection{Comparison of physical natures in Keplerian disks between B59\#11 and L1551NE}
\label{sec:dis:l1551}

Recently, a Keplerian disk was found in the proto-binary system L1551NE, which is a young Class-I protostar with a low bolometric temperature ($T_{\rm bol}$) of 91 K \citep{taka12}. B59\#11 also has a low bolometric temperature of 70 K, comparable to that of L1551NE. Thus, the comparison with L1551NE provides us with valuable information of a disk formation in the early phases of protostellar evolution. 
 
The dust mass associated with L1551NE obtained from the SMA observations is 0.016 $M_\sun$ \citep{taka12}. {This compares well} to that obtained for B59\#11 of 0.015 $M_\sun$ (see Section \ref{sec:dis:contdisk}) using the same mass opacity coefficient. The disk mass, central stellar mass, and the bolometric temperature of B59\#11 are comparable with those of L1551NE as shown in Table \ref{l1551ne}. The disk sizes {are possibly similar, too}. On the other hand, the disk around B59\#11 is a proto"stellar" disk, while the disk in L1551NE is a proto"binary" disk. 

\cite{machi04} have revealed that a disk is hard to fragment under a strong magnetic field even if the initial specific angular momentum is large. On the other hand, \citet{rom12} have suggested that a magnetic field with a strength of $\sim$0.1-0.2 mG is required to support the B59 clump against further fragmentation. This value seems to be larger than the strength of the magnetic field in the Taurus molecular cloud {($\sim$10-50 $\mu$G}; \citealt{lev01,cru03}). Thus, the strong magnetic field may keep B59\#11 {being a} single protostar.

Recent theoretical studies suggest that extended Keplerian disks may not be formed in the early phase of protostellar evolution, i.e., Class-0 phase {\citep[e.g.,][]{hen08,dap12,kura12}}. However, previous observations have shown that large Keplerian disks exist in the Class-I phase. In this paper, we have revealed that the Class-0/I protostar, B59\#11 has a large Keplerian disk with a size of 350 AU. Very recently, \cite{tob12} also discovered a large Keplerian disk around the Class-0 object, L1527. Hence, it remains unclear in what stage large Keplerian disks are formed. Further observational and theoretical studies are required to constraining the formation process of Keplerian disks in the protostellar evolution. 

\subsection{A rotating outflow in B59\#11?}
\label{sec:dis:out}

It is shown that there is a velocity gradient along the same direction as that of the envelope rotation in the B59\#11 outflow (Section \ref{sec:res:smaco}). There are some interpretations of this velocity gradient, and one of these is a rotating outflow. The rotations of the outflows in CO observations were reported in the Class-I and II sources (e.g., \citet{lau09,pech12}). The velocity structure in the northeastern lobe of the B59\#11 outflow is very similar to those in outflows ejected from  {the} Class-II YSO, CB26 \citep{lau09}, and {the} Class-I YSO, HH797 \citep{pech12}, which are two {very} reliable candidates {for a} rotating molecular outflow. The specific angular momentum of the outflow {in B59\#11} is estimated to be $\sim2.3\times10^{-3}$ km s$^{-1}$ pc from a separation of the peak positions of blueshifted and redshifted lobes of $\sim4$\arcsec (corresponding to 520 AU), a radial velocity of 1.8 km s$^{-1}$ (Figure \ref{smaout} and \ref{pv}), and the inclination angle of 75\arcdeg. It is noted that this value is almost the same as that of CB26 and smaller than that of HH797.

Another possibility is the outflow ejected from a protobinary system. Such a molecular outflow is identified in, for example, IRAS 05295+1247 \citep{arce05}, whose outflow is composed of one blueshited lobe and three redshifted lobes and may show a resemblance to the B59\#11 outflow. An outflow from a protobinary system, however, often shows a X-shaped structure, which comprises the walls of a cone-shaped outflow cavity, or the S-shaped structure, which is formed by a precessing outflow (e.g., \citet{arce05, wu09}). These structures are not confirmed in B59\#11 (Figure \ref{smaout}). 

B59\#11 might {constitute a} binary system with {the} nearby (projected separation is 1000 AU) protostar, B59\#11SW, as suggested by \cite{riaz09}. Assuming that the precession of the outflow axis is driven by tidal interactions between the circumstellar disk from which the jet is launched and a companion star on a non-coplanar orbit \citep[e.g.][]{lau09}, the precessing timescale is expected to be the same as the binary orbital period. If the stellar mass{es} of B59\#11 and B59\#11SW {are both close to} 1 $M_\sun$, the orbital period is estimated to be $\sim10^5$ yr assuming that the binary separation is 4000 AU ($\sim\mbox{1000 AU}/\cos{i}$, $i=75\arcdeg$) from the equation, $T^2=\frac{4\pi^2}{G(M_1+M_2)}a^3$ \footnote{$T$, $M_1+M_2$, and $a$ is a binary orbital period, masses of a primary and the secondary, and a semi major axis of the binary orbital, respectively.}. This period {would be much} longer than the outflow dynamical timescale of $\sim10^4$ yr \citep{dua12} obtained from their single-dish observations. Thus, the kinematics of the B59\#11 outflow can not be explained by the precession caused by such a very wide binary.

Is there really no {evidence} that B59\#11 is a closed binary system? If a binary system is formed, the tidal effects of the two protostars can transport rotational angular momentum of the common disk outwards and create a hole with a radius of $\sim$2.7$\times$(semi major axis of the binary orbital) (e.g., \citet{art91,dut94}). {Such} signature sometimes appears in, e.g., C$^{18}$O and/or $^{13}$CO images and P-V diagrams; continuous increasing of the rotation velocity as $r^{-1/2}$ and $r^{-1}$ is terminated at the inner edge of the disk. As discussed in Section \ref{sec:dis:linedisk}, the dense inner part traced by C$^{18}$O is likely to be a rotationally supported disk. And the high velocity emission in $^{13}$CO perhaps also traces the disk as shown in Figures \ref{pv} and \ref{velpos}. The highest velocity of the envelope/disk rotation is about 4.6 km s$^{-1}$ in $^{13}$CO, and the rotation velocity is estimated to trace $r$=30 AU region if we assume that $^{13}$CO emission traces the Keplerian disk at the inner area. The binary separation is calculated to be smaller than 11(=30/2.7) AU if we assume that the obtained radius is an upper limit to that of the inner edge of the disk and a binary companion exists. Further observations with higher spatial resolution are needed to confirm the outflow rotation in B59\#11.

\section{Conclusions}
\label{sec:summary}
We have carried out the ASTE observations toward the B59 region ($d\sim130$ pc) and a Class-0/I YSO, B59\#11 in the Pipe Nebula in 1.1 mm dust-continuum, CO(3--2), HCO$^+$, and H$^{13}$CO$^+$(4--3) emissions. We also processed archival data from the SMA observations of B59\#11 in 1.3 mm dust-continuum, CO, $^{13}$CO, and C$^{18}$O(2--1) emissions. The main results of these data are summarized as follows;

\begin{enumerate}

\item We have detected four dust condensations associated with YSOs, [BHB2007]\#1, \#9, \#10, and \#11 with the AzTEC/ASTE 1.1 mm continuum observations. The dust-continuum emission associated with B59\#11 is the strongest in the B59 region, and the mass of the condensation is estimated to be 0.09 $M_\sun$.

\item From ASTE CO(3--2) observations, we found that B59\#11 is blowing a collimated outflow whose axis is almost on the plane of the sky. This outflow {traces well} a cavity-like structure seen in the AzTEC/ASTE 1.1 mm dust-continuum map. The overall structures are in good agreement with the results by \citet{dua12}. 

\item The images of SMA $^{13}$CO and C$^{18}$O(2--1) emissions with a resolution of $\sim$5\arcsec\hspace{1ex}(corresponding to $\sim$650 AU) have revealed that a compact and elongated structure of the dense gas is associated with B59\#11 {with a size of} about 4000$\times$2000 AU (with an aspect ratio of $\sim$2:1). The dense gas shows a rotation along its major axis and the specific angular momentum is estimated to be 2.1$\times10^{-3}$ km s$^{-1}$ pc. ASTE HCO$^+$ emission also shows a velocity gradient, which is considered to be arisen from the large-scale envelope rotation.

\item A compact dust-continuum condensation with a mass of 7.3$\times10^{-2}$ $M_\sun$ is identified from our SMA data in 1.3 mm continuum emission. The deconvolved size of the dust condensation is estimated to be $\sim$350$\times$180 AU, and is oriented along the rotation axis of the dense gas. 

\item The SMA CO(2--1) emission {map} shows a velocity gradient in the outflow lobe. It is considered to trace the outflow rotation and its specific angular momentum is estimated to be 2.3$\times10^{-3}$ km s$^{-1}$ pc. This specific angular momentum is comparable to that of another rotating outflow, CB26.  Further observations with higher spatial resolution, however, are needed to confirm the rotation of the B59\#11 outflow.  

\item The radial velocities in SMA $^{13}$CO and C$^{18}$O(2--1) emissions have different power-law indexes, and estimated to be -1.3 and -0.61, respectively, implying that C$^{18}$O emission traces a Keplerian disk. The existence of the disk is also suggested from our analysis of ASTE and SMA data in dust-continuum emissions. The disk mass and the central stellar mass are estimated to be 0.03 and 0.73 $M_\sun$, respectively.

\end{enumerate}

\clearpage

\bibliography{sma_compact}

\clearpage

\begin{deluxetable}{lcccc}
\tabletypesize{\small}
\tablewidth{0pt}
\tablecolumns{4}
\tabletypesize{\scriptsize}
\rotate
\tablecaption{Parameters for the ASTE observations}
\tablehead{
\colhead{Telescope/Receiver} & \colhead{ASTE/AzTEC} & \colhead{ASTE/CATS345} & \colhead{ASTE/CATS345} & \colhead{ASTE/CATS345}  
}
\startdata
Line/Frequency/Wavelength & 1.1 mm & $^{12}$CO($J$=3--2; 345.796 GHz) & HCO$^+$($J$=4--3; 356.734 GHz) &H$^{13}$CO$^+$($J$=4--3; 346.998 GHz)\\
 Observation date& 2008 Oct 17 - 31 & 2011 May 30 -  Jun 1   & 2011 May 30 -  Jun 1, 2012 January 23 & 2012 January 23\\
 Observing mode & Raster & OTF/Position switch & Position Switch & Position Switch\\
 Mapping size & 35\arcmin$\times$35\arcmin & 15\arcmin$\times$11\arcmin \& 80\arcsec$\times$80\arcsec & 80\arcsec$\times$80\arcsec \& 60\arcsec$\times$60\arcsec & 60\arcsec$\times$60\arcsec\\
Effective beam size & 36\arcsec & 27\arcsec (OTF) / 22\arcsec (Position Switch) & 22\arcsec &22\arcsec\\
 Velocity resolution & --  & 0.10 km s$^{-1}$ & 0.11 km s$^{-1}$ & 0.10 km s$^{-1}$\\
 Typical rms in $T_{\rm A}^*$ & 7 mJy beam$^{-1}$ & 0.36 K & 0.1 K \& 0.03 K & 0.03 K \\
\enddata
\label{ASTE observations}
\end{deluxetable}

\clearpage
\begin{deluxetable}{lcccc}
\tabletypesize{\scriptsize}
\tablewidth{0pt}
\tablecolumns{5}
\tablecaption{Parameters for the SMA data reductions}
\tablehead{
\colhead{Line/Wavelength} & \colhead{1.3 mm} & \colhead{$^{12}$CO($J$=2--1)} & \colhead{$^{13}$CO($J$=2--1)}  & \colhead{C$^{18}$O($J$=2--1)}
}
\startdata
Frequency(GHz) & 220.5 \& 230.5 & 230.538 & 220.399 & 219.560 \\
Observation date & \multicolumn{4}{c}{2008 Mar 25} \\
Array Configuration & \multicolumn{4}{c}{Compact (7 ant, Minimum Baseline=7 k$\lambda$, Maximum Baseline=50 k$\lambda$)} \\
Bandwidth / Channel Separation & 2.0 + 2.0 GHz & 1.1 km s$^{-1}$ & 1.1 km s$^{-1}$ & 1.1 km s$^{-1}$ \\
Pointing Center & \multicolumn{4}{c}{$(\alpha_{\rm J2000}, \delta_{\rm J2000}$)=($17^{\rm h}11^{\rm m}23\fs18, -27\arcdeg24\arcmin31\farcs5$)}\\
On Source Time &  \multicolumn{4}{c}{14 min} \\
System Temperature &  \multicolumn{4}{c}{80 - 130 K in SSB} \\
Bandpass Calibrators &  \multicolumn{4}{c}{3C454.3} \\
Complex Gain Calibrator &  \multicolumn{4}{c}{NRAO530, J1924-292}\\
Absolute Flux Calibrators & \multicolumn{4}{c}{Callisto}\\
Beam Size & 5\farcs0$\times$2\farcs8 (31) & 4\farcs8$\times$2\farcs8 (31) & 4\farcs8$\times$2\farcs9 (30) & 5\farcs2$\times$2\farcs9 (32)\\
map rms & 12 mJy beam$^{-1}$ & 200 mJy beam$^{-1}$ & 200 mJy beam$^{-1}$ & 200 mJy beam$^{-1}$\\
\enddata
\label{SMAobs}
\end{deluxetable}

\clearpage
\begin{deluxetable}{lcccccccccc}
\tablewidth{0pt}
\tablecolumns{11}
\tabletypesize{\scriptsize}
\rotate
\tablecaption{Parameters of AzTEC/ASTE 1.1 mm dust-continuum condensations associated with YSOs}
\tablehead{
\colhead{ID$^{\rm a}$ } & \colhead{Spectrum$^{\rm a}$} & \colhead{${\alpha_{\rm J2000}}$}& \colhead{${\beta_{\rm J2000}}$}& \colhead{Peak Flux} & \colhead{Integrated Flux } & \colhead{Source Size} &\colhead{Beam Deconvolve} & \colhead{Mass$^{\rm b}$} & \colhead{Density} &\colhead{Separation from}
\\
\colhead{} &\colhead{Type}& \colhead{} & \colhead{} & \colhead{Density [Jy beam$^{-1}$]} & \colhead{Density [Jy]} & \colhead{ [\arcsec$\times$\arcsec (P.A.\arcdeg)]} & \colhead{ Size [\arcsec$\times$\arcsec]
(P.A.\arcdeg)]} & \colhead{ [M$_{\sun}$]} & \colhead{ [$\times10^6$cm$^{-3}$]} & \colhead{Protostars$^{\rm e}$ [\arcsec]}}
\startdata
1$^{\rm c}$ &Flat & 17:11:04.08 & -27:22:57.0 & 0.26 & 0.26 & 36$\times$34(58) & \nodata$^{\rm c}$ & 0.37 & 0.98$^{\rm d}$ &  3.1\\
3$^{\rm d}$  & I\hspace{-0.5ex}I & 17:11:12.46 & -27:27: 00.1 & 0.07 & 0.2 & 73$\times$15(-28) & 64$\times$52 & 0.35 & 0.20 & 11 \\
7$^{\rm d}$ & Flat & 17:11:17.60 & -27:25:16.2 & 0.11 & 0.2 & 60$\times$44(0) & 48$\times$27 & 0.34 & 0.81 & 9.3\\
9 & Flat & 17:11:21.52 & -27:27:40.5 & 0.28 & 0.3 & 46$\times$42(-14) & \nodata$^{\rm c}$ & 0.43 &  1.0$^{\rm d}$ & 1.8\\
10 & I & 17:11:21.50 & -27:26:11.7 & 0.44 & 1.1 & 76$\times$51(55) & 63$\times$25 & 1.6 & 2.9 & 13 \\
11 & 0/I & 17:11:23.22 & -27:24:35.6 & 1.3 & 1.3 & 41$\times$38(-63) & \nodata$^{\rm c}$ & 0.09$^{\rm f}$ & 1.9$^{\rm g}$ & 4.2\\
14 & I\hspace{-0.5ex}I & 17:11:25.21 & -27:25:36.2 & 0.4 & 1.8 & 110$\times$52(14) & 103$\times$39 & 2.5 &  0.65 & 29 \\
(SMA) & & & & & & & & & \\
11$^{\rm e}$ & 0/I & 17 11 23.08 & -27 24 33.8 & 0.43 & 0.60 & 5.56$\times$3.49 (22) & 2.8$\times$1.4 (-11) & 0.015 & 2.7$\times 10^{\rm 3\hspace{1ex}g}$
\enddata
\tablenotetext{a}{From \citet{bro07}.}
\tablenotetext{b}{Estimated on the assumption of $T_{\rm d}$=10 K \citep{rath08}, $\kappa_\nu=0.1(250\micron/\lambda)^{\beta}$ cm$^2$ g$^{-1}$ \citep{hil83}, and $\beta$=2.}
\tablenotetext{c}{These dust-continuum sources are strongly suggested to be unresolved, and we cannot obtain the deconvolved sizes.}
\tablenotetext{d}{\hspace{1ex}\hspace{-1ex}Lower limits obtained using {a} beam size of 36\arcsec. }
\tablenotetext{e}{Separations between the peak positions of AzTEC/ASTE 1.1 mm dust-continuum condensations and protostars identified by \citet{bro07}. }
\tablenotetext{f}{Estimated on the assumption of $T_{\rm d}$=30 K and $\beta$=1.0. See Section \ref{sec:res:aztc} in the text. }
\tablenotetext{g}{Estimated using the deconvolved size of 18\arcsec$\times$17\arcsec obtained by \citet{rom12} from MAMBO-II 1.2 mm observations. }
\label{AzTEC sources}
\end{deluxetable}

\clearpage
\begin{deluxetable}{lcccccc}
\rotate
\tablewidth{0pt}
\tablecolumns{7}
\tablecaption{Outflow Parameters}
\tabletypesize{\scriptsize}
\tablehead{
\colhead{ } & \colhead{\hspace{1ex}} & \colhead{Velocity [km s$^{-1}$] $^{\rm a}$} &\colhead{Total Integrated Intensity}& \colhead{Mass [M$_{\sun}$]$^{\rm b}$} & \colhead{Energy [M$_{\sun}\cdot$ (km s$^{-1})^2$]$^{\rm c}$} & \colhead{Momentum [M$_{\sun} \cdot$ (km s$^{-1}$)]$^{\rm c}$}}
\startdata
ASTE & blue & -2.5$\sim$1.5 & 55 K$\cdot$km s$^{-1}$& 9.5$\times 10^{-3}$ & 1.3$\times10^{-1}$ & 0.89 \\
 & red & 5.5$\sim$9.5 & 120 K$\cdot$km s$^{-1}$&2.0$\times 10^{-2}$ & 2.1$\times10^{-1}$ & 1.2 \\
SMA & blue & -5.0$\sim$1.5 & 36 Jy$\cdot$km s$^{-1}$ & $<$2.8$\times10^{-4}$ & $<$4.0$\times 10^{-3}$ & $<$3.0$\times10^{-2}$ \\
 & red & 4.5$\sim$12 & 67 Jy$\cdot$km s$^{-1}$ & $<$4.4$\times 10^{-4}$ & $<$3.4$\times 10^{-3}$ & $<$1.6$\times10^{-2}$
\enddata
\tablenotetext{a}{The velocity ranges where emissions are detected above the 3$\sigma$ level.}
\tablenotetext{b}{Masses are estimated on the assumption of optically thin emission, $T_{\rm ex}$=10 K, and $X[{\rm CO}]=10^{-4}$.}
\tablenotetext{c}{Values are estimated on the assumption of $i$=75\arcdeg\hspace{1ex}and $V_{\rm sys}=3.6$ km s$^{-1}$.}
\label{COpar}
\end{deluxetable}

\clearpage
\begin{deluxetable}{lcccccc}
\tablewidth{0pt}
\tablecolumns{5}
\tablecaption{Masses in HCO$^+$ and H$^{13}$CO$^+$(4--3) emissions}
\tabletypesize{\scriptsize}
\tablehead{
\colhead{Line} & {} & {-1.5$\sim$2 km s$^{-1}$} & {2$\sim$4 km s$^{-1}$} & {5$\sim$8 km s$^{-1}$} & {Total}
}
\startdata
HCO$^+$(4--3) & integrated intensity (K($T_{\rm A}^\ast$)$\cdot$km s$^{-1}$) & 0.28 & 2.5 & 0.41 & 3.2\\
 & mass ($M_\sun$)$^{\rm a}$ & 1.2$\times10^{-2}$ &\nodata & 1.8$\times10^{-2}$ & 3.0$\times10^{-2}$ \\
H$^{13}$CO$^+$(4--3) & integrated intensity (K($T_{\rm A}^\ast$)$\cdot$km s$^{-1}$) & \nodata & 0.17 & \nodata &  \\
 & mass ($M_\sun$)$^{\rm a}$ & \nodata & 0.53 &\nodata & 0.53
\enddata
\tablenotetext{a}{Masses are estimated on the assumption of optically thin emission, $T_{\rm ex}=30$ K, $X[{\rm HCO^+}]=1.0\times10^{-10}$, and $X[{\rm H^{13}CO^+}]=1.4\times10^{-12}$.}
\label{HCO+par}
\end{deluxetable}

\clearpage
\begin{deluxetable}{ccccc}
\tablecolumns{4}
\tabletypesize{\scriptsize}
\tablecaption{Results of $\chi^2$-fitting}
\tablehead{
\colhead
{} & {Coefficient (a)$^{\rm a}$} & {Power-law index (b) $^{\rm a}$}  & {Systemic velocity ($v_{\rm sys}$) $^{\rm a}$} & {$\sqrt{\chi^2}$}
}
\startdata
$^{13}$CO(2--1)  & 4.4 & -1.3 & 3.3 & 1.1 \\
C$^{18}$O(2--1)  & 2.5 & -0.61 & 3.3 & 1.1$\times10^{-1}$ \\
C$^{18}$O(2--1)  & 2.2  & -0.5 & 3.3 & 6.8$\times10^{-1}$ \\
(fixed on Keplerian rotation) & 0.73 $M_\sun$ $^{\rm b}$& & &
\enddata
\label{fit}
\tablenotetext{a}{$|v_{\rm LSR}-v_{\rm sys}|=a\times r^b$ and $r$ is the separation of a peak position with the 1.3 mm dust-continuum peak.}
\tablenotetext{b}{The central stellar mass estimated from the fitting with a Keplerian rotation curve assuming the inclination angle of 75\arcdeg. }
\end{deluxetable}

\clearpage
\begin{deluxetable}{ccc}
\tablewidth{0pt}
\tablecolumns{3}
\tablecaption{Results of SED fiting}
\tablehead{
\colhead{Parameter} & {Edge-on} & {\citet{riaz09}} 
}
\startdata
$R_*$ ($R_\sun$) & 5.6 & 3.4 \\
$T_*$ (K) & 4000 & 3300  \\
$M_*$ ($M_\sun$) & 0.81 & 0.28 \\
$\theta_{\rm in}$ & 75\arcdeg & 53\arcdeg-59\arcdeg \\
$R_{\rm env,max}$ (AU) & 3000 & 4000 \\
$R_{\rm disk,max}$ (AU) & 390 & 30 \\
$M_{\rm disk}$ ($M_\sun$) &  7.0$\times10^{-3}$ & 2.6$\times10^{-2}$
\enddata
\label{sedpar}
\end{deluxetable} 

\clearpage
\begin{deluxetable}{ccc}
\small
\tablewidth{0pt}
\tablecolumns{3}
\tablecaption{Comparison with L1551NE}
\tablehead{
\colhead{Parameter} & {B59\#11} & {L1551NE} 
}
\startdata
$L_{\rm bol}$ ($L_\sun$) & 2.0 & 4.2 \\
$T_{\rm bol}$ (K) & 70 & 91 \\
$R_{\rm disk}$ (AU) & $<$350 & 300 \\
$M_*$ ($M_\sun$) & 0.84 & 0.8 \\
$M_{\rm disk}$ ($M_\sun$)& 0.016 &  0.015 \\
$R_{\rm env}$ (AU)&  2300 $^{\rm a}$  & $>$8500 $^{\rm b}$  \\
$M_{\rm env}$ ($M_\sun$) & 0.082 & 0.24 \\
specific angular momentum (km s$^{-1}$ pc) & 1.9$\times 10^{-3}$ & 6$\times 10^{-4}$ \\
\hline
\enddata
\scriptsize
\tablenotetext{a}{The deconvolved size obtained from IRAM/MAMBO-II 1.2 mm continuum observations \citep{rom12}}
\tablenotetext{b}{The size of the sharp edge of the density profile obtained from IRAM/MPIfR 1.3 mm continuum observations \citep{mot01}}
\label{l1551ne}
\end{deluxetable}

\clearpage

\clearpage

\clearpage

\clearpage
\begin{figure} 
\epsscale{1.0}
\plotone{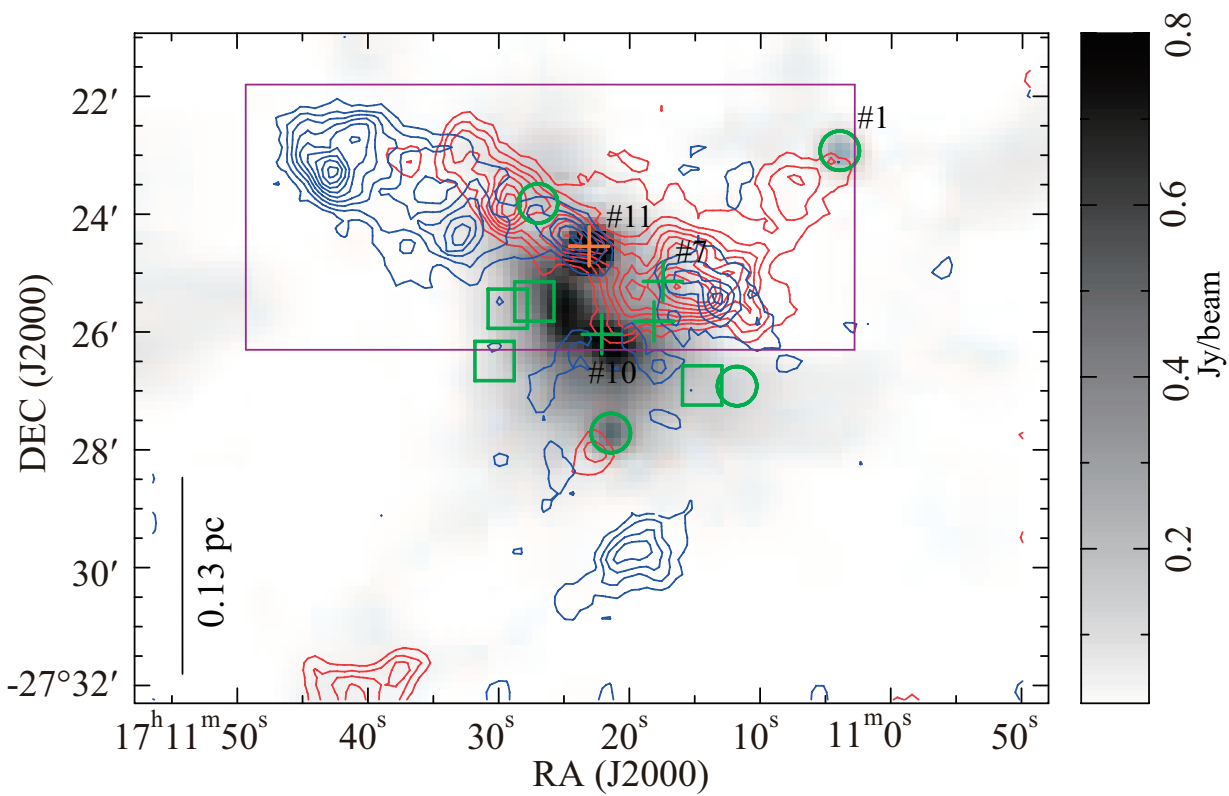}
\caption{ASTE CO(3--2) map of the B59 overlaid on the AzTEC/ASTE 1.1 mm dust-continuum image. The blue and red contours represent blueshifted and redshifted CO(3--2) intensities integrated between $V_{\rm LSR}=-2.5$ to 1.5 km s$^{-1}$ and 5.5 to 9.5 km s$^{-1}$, respectively. Contour levels start at 3$\sigma$ noise level with an interval of 2$\sigma$ (1$\sigma$=0.3 K($T_{\rm A}^*$)$\cdot$ km s$^{-1}$). The symbols are the same as those in Figure \ref{aztec}. The purple box shows the area of Figure \ref{COch}.}
\label{COwing}
\end{figure}

\clearpage
\begin{figure} 
\epsscale{0.8}
\plotone{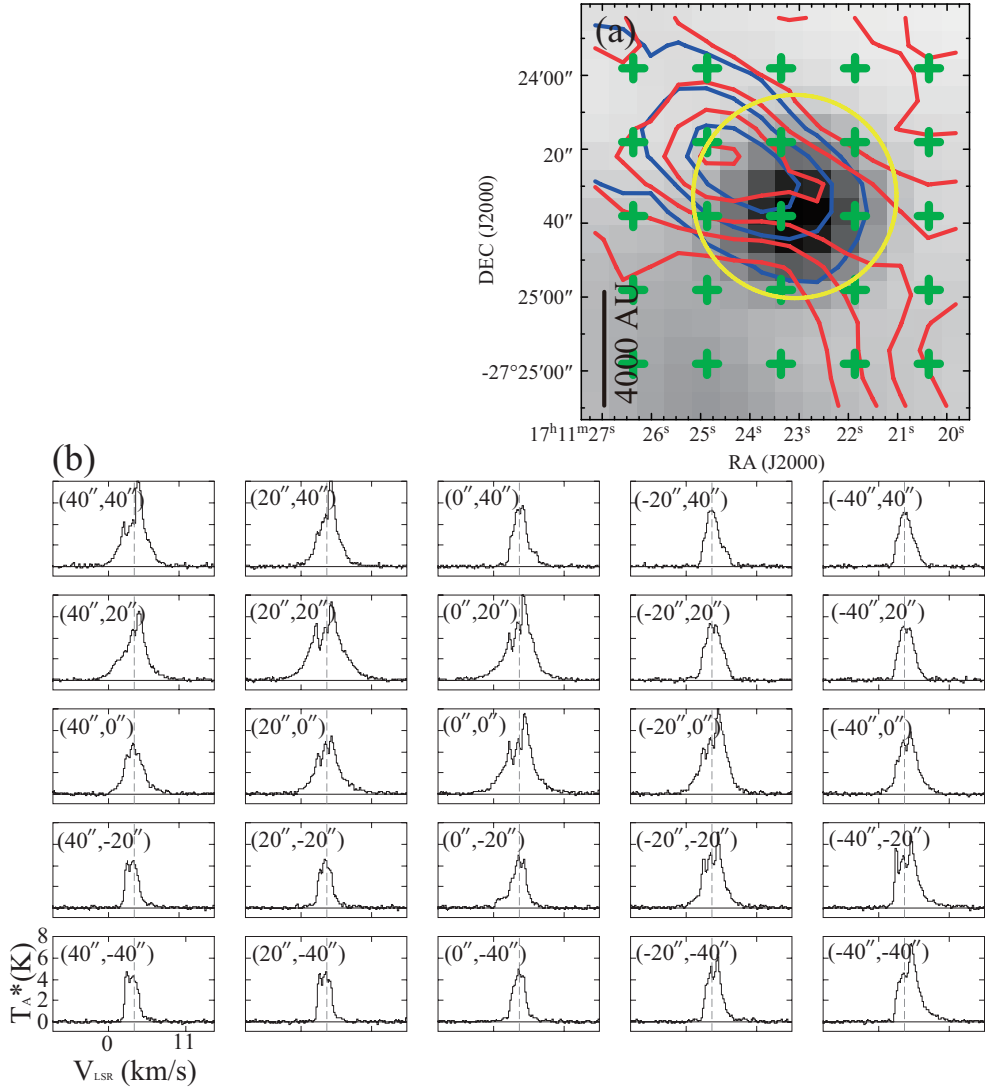}
\caption{(a) CO(3--2) map (the same as Figure \ref{COwing}), and (b) CO(3--2) profile map taken toward B59\#11. The green crosses in Figure (a) show the positions where profiles in Figure (b) were obtained, and the yellow circle shows the SMA FoV. A grid spacing of Figure (b) is 20\arcsec. The dotted vertical lines are the systemic velocity of 3.6 km s$^{-1}$ obtained from ASTE H$^{13}$CO$^+$(4--3) observations (Figure \ref{HCO+pro}). The (0\arcsec,0\arcsec) position is the peak position of AzTEC/ASTE 1.1 mm dust-continuum condensation associated with B59\#11, ($\alpha_{J2000.0}$,$\delta_{J2000.0}$)=(17$^{\rm h}$11$^{\rm m}$23\fs1 , -27\arcdeg24\arcmin38\farcs2).}
\label{COpro}
\end{figure}

\clearpage
\begin{figure} 
\epsscale{1.0}
\plotone{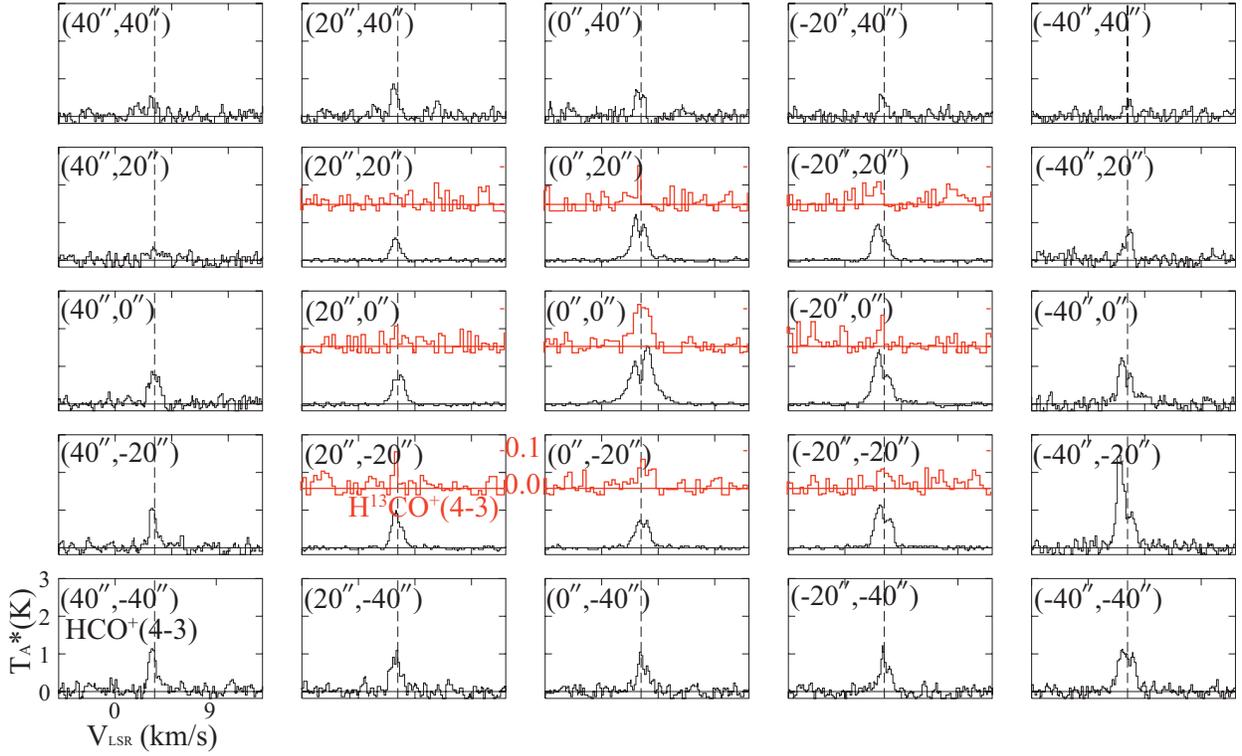}
\caption{ASTE HCO$^+$(4--3) (black lines) and H$^{13}$CO$^+$(4--3) (red lines) profile maps. The dotted vertical lines are the systemic velocity. The (0\arcsec,0\arcsec) position and a grid spacing is the same as those of Figure \ref{COpro} (b). }
\label{HCO+pro}
\end{figure}

\clearpage
\begin{figure} 
\epsscale{1.0}
\plotone{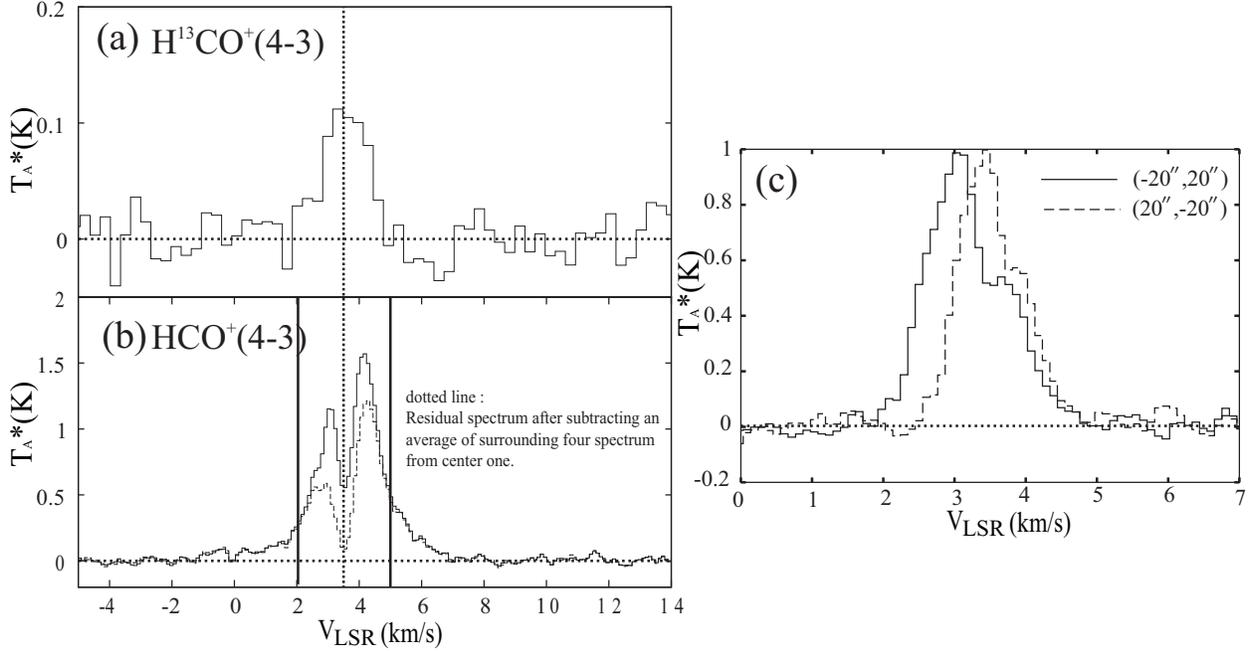}
\caption{The profiles of (a) H$^{13}$CO$^+$(4--3) and (b) HCO$^+$(4--3) lines at the (0\arcsec,0\arcsec) position in Figure \ref{HCO+pro}. The dotted line in Figure (b) shows the residual spectrum after subtracting an average of surrounding four spectra from that at the center. The dotted vertical lines are the systemic velocity and the solid vertical lines in (b) indicate the velocity range where the masses of the high velocity components are obtained (Table \ref{HCO+par}). (c) Line profiles of HCO$^+$(4--3) at the positions of (-20\arcsec,20\arcsec) and (20\arcsec,-20\arcsec). }
\label{HCO+rat}
\end{figure}

\clearpage
\vspace{-20ex}

\clearpage

\clearpage
\begin{figure} 
\epsscale{1.0}
\plotone{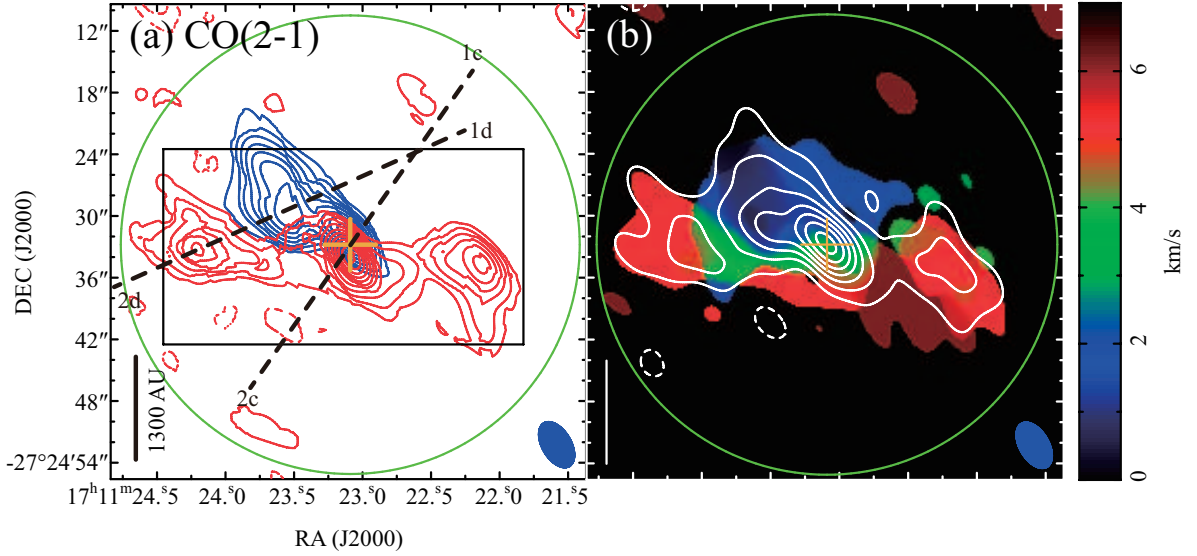}
\caption{The maps of SMA CO(2--1) emission. The blue and red contours represent blueshifted and redshifted intensities integrated between $V_{\rm LSR}$=-4.9 to 1.4 km s$^{-1}$ and 4.6 to 12 km s$^{-1}$, respectively. The white contours superposed on the mean velocity  map in (b) indicate the emission integrated over intervals from $V_{\rm LSR}$=-4.9 to 12 km s$^{-1}$. Contour levels start at 3$\sigma$ noise level with intervals of (a) 1.5$\sigma$ (1$\sigma$=1.0 Jy beam$^{-1}$$\cdot$km s$^{-1}$), and (b) 3$\sigma$ (1$\sigma$=1.8 Jy beam$^{-1}$$\cdot$km s$^{-1}$). The symbols in each figure are the same as those in Figure \ref{C11}. The dashed lines in (a) are cuts along which the P-V diagrams in Figure \ref{pv} (c) and (d) were produced.}
\label{smaout}
\end{figure}

\clearpage
\vspace{-15ex}
\begin{figure} 
\epsscale{1.0}
\plotone{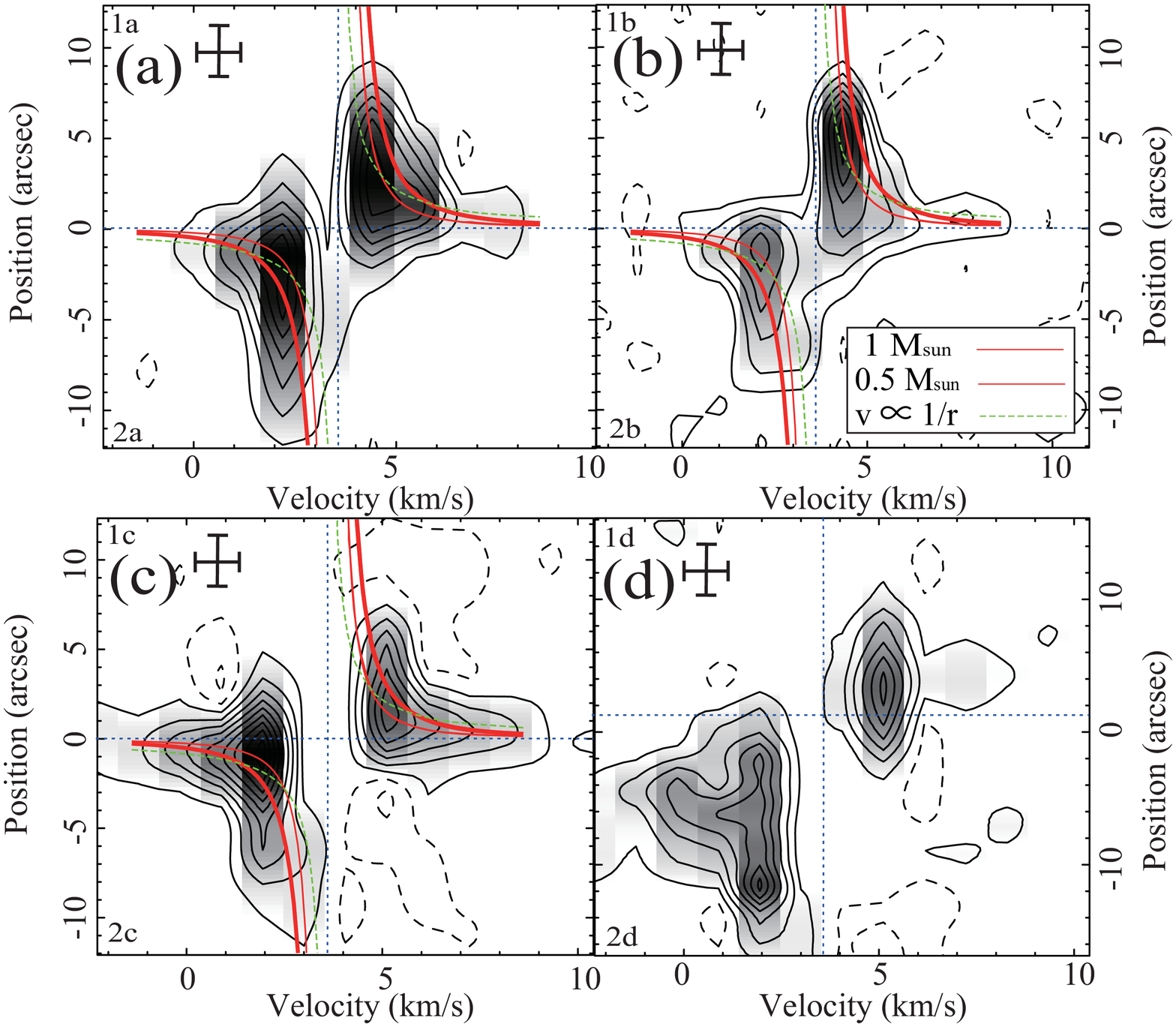}
\caption{Position-Velocity diagrams in (a) SMA $^{13}$CO(2--1), (b) C$^{18}$O(2--1), and (c) (d) CO(2--1) emissions, cut along lines with a PA of 145\arcdeg \hspace{1ex}centered at B59\#11 (Figure \ref{smadense}, \ref{smaout}). Contour levels are start at (a) (c) (d) 3$\sigma$ and (b) 1.5$\sigma$ noise level with intervals of (a) 3$\sigma$ (1$\sigma$=0.2 Jy beam$^{-1}$), (b) 1.5$\sigma$ (1$\sigma$=0.2 Jy beam$^{-1}$), and (c) (d) 6$\sigma$ (1$\sigma$=0.2 Jy beam$^{-1}$), respectively. The blue dashed lines in the vertical and horizontal directions show the systemic velocity and the position of B59\#11. The red thick and narrow lines show Kepelerian disk models for the central masses of 1.0 $M_\sun$ and 0.5 $M_\sun$, and the green curves show the rotation conserving its angular momentum ($r\propto v^{-1}$) whose rotation velocity is 2 km s$^{-1}$ at a radius of 200 AU.}
\label{pv}
\end{figure}

\clearpage
\end{document}